# Molecular Orbital Projectors in Non-empirical jmDFT Recover Exact Conditions in Transition Metal Chemistry

Akash Bajaj[1,2], Chenru Duan[1,3], Aditya Nandy[1,3], Michael G. Taylor[1], and Heather J. Kulik[1,*]

[1]Department of Chemical Engineering, Massachusetts Institute of Technology, Cambridge, MA 02139

[2]Department of Materials Science and Engineering, Massachusetts Institute of Technology, Cambridge, MA 02139

[3]Department of Chemistry, Massachusetts Institute of Technology, Cambridge, MA 02139

ABSTRACT: Low-cost, non-empirical corrections to semi-local density functional theory are essential for accurately modeling transition metal chemistry. Here, we demonstrate the judiciously-modified density functional theory (jmDFT) approach with non-empirical U and J parameters obtained directly from frontier orbital energetics on a series of transition metal complexes. We curate a set of nine representative Ti(III) and V(IV) $d^1$ transition metal complexes and evaluate their flat plane errors along the fractional spin and charge lines. We demonstrate that while jmDFT improves upon both DFT+U and semi-local DFT with the standard atomic orbital projectors (AOPs), it does so inefficiently. We rationalize these inefficiencies by quantifying hybridization in the relevant frontier orbitals for both the case of fractional spins and fractional charges. To overcome these limitations, we introduce a procedure for computing a molecular orbital projector (MOP) basis for use with jmDFT. We demonstrate this single set of $d^1$ MOPs to be suitable for nearly eliminating all energetic delocalization error and static correlation error. In all cases, the MOP jmDFT outperforms AOP jmDFT, and it eliminates most flat plane errors at non-empirical values. Unlike widely employed DFT+U or hybrid functionals, jmDFT nearly eliminates energetic delocalization error and static correlation error within a non-empirical framework.

# 1. Introduction.

Applied Kohn–Sham density functional theory (DFT)[1-4] is widely employed in transition metal chemistry due to its good balance of cost and accuracy[5,6]. However, common local[7,8] and semi-local[9-11] approximations to the exchange-correlation (xc) functional within DFT exhibit one- and many-electron self-interaction errors (SIEs) also referred to as delocalization error (DE).[12-16] These errors can be traced back to the fact that these approximations lack key properties of the exact functional, including a derivative discontinuity and piecewise linearity upon fractional charge addition[17-23]. These DEs[24-27] are associated with erroneous predictions of densities[24,28-32], electron affinities[33-36], band gaps[27,37,38], spin-state ordering[39-47] and other properties[48-51]. Among the possible generalizations[52,53] to Kohn–Sham DFT that help mitigate these errors[51,54-66], the DFT+U approach[67-71] is commonly employed for transition-metal-containing systems[39,72-83] as it allows a targeted accounting of electronic correlation of the localized electrons (i.e., $d$ or $f$) at the cost of semi-local DFT[70].

The simplified[69,71], rotationally invariant[68] form of DFT+$U$ is:

$$E^{\mathrm{DFT+U}} = E^{\mathrm{DFT}} + \frac{1}{2}\sum_{I,\sigma}\sum_{nl} U^I_{nl}\left[\mathrm{Tr}\left(\mathbf{n}^{I\sigma}_{nl}\left(\mathbf{1}-\mathbf{n}^{I\sigma}_{nl}\right)\right)\right] . \qquad (1)$$

where a +U correction is applied on every $nl$ subshell and spin $\sigma$ on atom $I$. The occupation matrix $\mathbf{n}^{I\sigma}_{nl}$ is obtained by projecting states $|\psi_{k,v}\rangle$ onto localized atomic orbitals (AOs) from the $nl$ subshell with angular momenta $m$ $|\phi^I_m\rangle$ on site $I$ using:

$$n^{I\sigma}_{mm'} = \sum_{k,v}\langle\psi_{k,v}|\phi^I_{m'}\rangle\langle\phi^I_m|\psi_{k,v}\rangle . \qquad (2)$$

The deviation from linearity of the energy[17] (i.e., the energetic delocalization error, EDE[28]) is approximately quadratic for most exchange-correlation (xc) functionals[18,84]. Thus, the first-principles motivation for using DFT+U to correct semi-local DFT errors is made clear by the fact that the quadratic DFT+U expression in eq. (1) should recover piecewise linearity[71]. The only caveat to this analysis is that we require the fractional electron addition to be captured by the projection onto AOs in $\mathbf{n}_{nl}^{I\sigma}$.[84] For the simplifying case where electron addition produces quadratic energy dependence and is localized to a single element of the occupation matrix, the $U$ value that recovers piecewise linearity[84-86] is the constant energetic curvature[84] of the original xc functional.

The standard DFT+U approach[67,71,87,88] applied on the metal-centered $d$ atomic orbitals (AOs) of transition metal (TM) complexes is known to reduce their EDE, albeit at rates lower than the maximum 0.125 eV/eV expected at the atomic limit[84,89]. Moreover, DFT+U can eliminate the EDE entirely at the non-empirically determined $U$ (i.e., the global curvature[18,84]):

$$U = \langle \frac{\partial^2 E}{\partial q^2} \rangle = \varepsilon_{N+1}^{\text{HOMO}} - \varepsilon_{N}^{\text{LUMO}}, \qquad (3)$$

where the global curvature can be suitably approximated to be constant and estimated as the difference in the $N+1$–electron highest occupied molecular orbital (HOMO) and $N$-electron lowest unoccupied molecular orbital (LUMO) eigenvalues along the fractional charge line (FCL). Standard DFT+U can fail completely and have no effect on EDE for pathological cases[84,90], which can typically be attributed to strong or variable hybridization of MOs. To overcome these limitations, we introduced[89] a simple scheme for obtaining molecular orbital projectors (MOPs) in DFT+U to eliminate EDE at non-empirical $U$ values. Similar approaches have been useful in the solid state[91,92]. Nevertheless, DFT+U worsens the SCE for all systems[85]. Given the reliance of many high-throughput screening workflows on DFT+U to improve properties without increasing cost[93], a robust low-cost alternative is needed.

Toward this end, we introduced[85,86] the judiciously-modified DFT (jmDFT) approach, wherein the shape of errors can be fit to any analytical function and then inverted to correct errors along the flat plane. A similar idea has been recently demonstrated with deep learning[94]. The approach also bears some similarities to local orbital scaling corrections.[95,96]

The jmDFT approach with atomic projections has been shown as one path to eliminate the EDE and SCE simultaneously for isolated atoms and diatomics[85]:

$$E^{\text{DFT+U+J}} = E^{\text{DFT}} + \frac{1}{2}\sum_{I,\sigma}\sum_{nl}(U_{nl}^{I} - J_{nl}^{I})\left[\text{Tr}\left(\mathbf{n}_{nl}^{I,\sigma}\left(\mathbf{1} - \mathbf{n}_{nl}^{I,\sigma}\right)\right)\right] + \frac{1}{2}\sum_{I,\sigma}\sum_{nl}J_{nl}^{I}\left[\text{Tr}\left(\mathbf{n}_{nl}^{I,\sigma}\mathbf{n}_{nl}^{I,-\sigma}\right)\right] \quad (4)$$

where the first term in this equation corresponds to eqn. **Error! Reference source not found.** except that $U$ has been replaced by $U$-$J$. The $J$ term alone is insufficient to eliminate energetic errors across the full flat plane due to its monotonic dependence on electron count.[85] For $s$-electron valences we introduced a modified $J$ (i.e., $J'$) term:

$$E^{J'} = \frac{1}{2}\sum_{I,\sigma}\sum_{nl}J'^{I}_{nl}\left[\text{Tr}\left(\left(\mathbf{n}_{nl}^{I,\sigma} - 1\right)\left(\mathbf{n}_{nl}^{I,-\sigma} - 1\right)\right)\right] \quad (5)$$

The coefficients can all be determined non-empirically[86] by extracting the parameter values (i.e., for $U$, $J$ or $J'$) from the DFA needing correction. In this case, the $U$ for jmDFT is obtained using the global curvature estimate of eq. (3). The $J$ (or $J'$) is obtained using $U + J \sim \varepsilon_{N+1}^{\text{HOMO}} - \varepsilon_{N+1}^{\text{LUMO}}$, where the N+1–electron system corresponds to one endpoint of the seam along the flat plane. However, the suitability of this jmDFT approach with standard atomic projections on the EDE and SCE of TM complexes has not been demonstrated.

Analysis of the exact flat-plane constraint[97] is typically represented as the variation in energies as electrons get added from a singlet $N - 1$ electron system to another singlet $N + 1$ electron system passing through the $N$-electron doublet configurations (Supporting Information Figure S1). To demonstrate removal of flat plane errors within TM complexes, a straightforward case where electron addition leads from a singlet to a doublet state is when the metal center has a formal $d^0$ and $d^1$ configuration respectively at the atomic limit. Thus, in this work we investigate the suitability of both standard (i.e., AOP) and molecular (i.e., MOP) jmDFT with non-empirical coefficients to recover exact conditions for $d^1$ Ti(III) and V(IV)-based complexes (Figure 1 and Supporting Information Text S1 and Table S1, also see Computational Details). We demonstrate that jmDFT never worsens static correlation error when improving delocalization error and that the MOP approach is essential for capturing the degree of hybridization present in the frontier orbitals of these complexes.

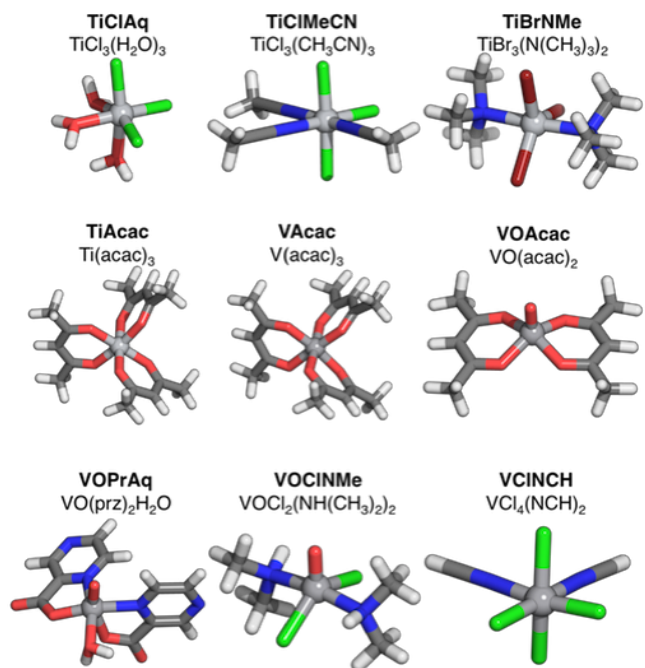

**Figure 1.** DFT optimized structures of Ti and V mononuclear complexes used for flat-plane error analysis. Chemical formula of each structure is annotated, and the nomenclature used for referencing each structure throughout this work is indicated in bold at the top of each structure.

Metal centers are shown as silver spheres whereas hydrogen, carbon, nitrogen, oxygen, chlorine and bromine atoms are shown in white, gray, blue, red, green and brown respectively.

## 2. Computational Details.

We obtained all mononuclear Ti/V transition metal (TM) complexes with a formal $d^1$ electron configuration using the Cambridge Structural Database (CSD)[98] v5.41 (with August 2020 update). From a total of 377 initial candidate structures, we selected 50 representative Ti(III)- and V(IV)-based TM complexes, along with an additional V(acac)$_3$ complex in the +3 oxidation state after validating their user-level metal oxidation state assignment (Supporting Information Text S1). Geometry optimizations of all 50 structures were carried out using TeraChem[99] v1.9 using the hybrid generalized gradient approximation (GGA)[11] PBE0 exchange-correlation (xc) functional[100,101] and the LACVP* composite basis set, which consists of a LANL2DZ effective core potential[102,103] for transition metals, Br and I and the 6-31G* basis[104] for all other atoms. Geometry optimizations were carried out in translation rotation internal coordinates[105] using the L-BFGS algorithm[106-111] with default thresholds of $4.5\times10^{-4}$ Hartree/bohr for the gradient and $1\times10^{-6}$ Hartree for changes in the self-consistent field (SCF) energy. From these optimized structures, we first selected seven representative cases that were validated to have high fidelity (i.e., root mean squared deviation i.e., RMSD < 0.4 Å) with the CSD structure (Supporting Information Table S1). Two Ti complexes (i.e., TiCl$_3$(H$_2$O)$_3$ and TiCl$_3$(CH$_3$CN)$_3$) required separate optimizations with an implicitly modeled solvent using the conductor-like polarizable continuum model (C-PCM)[112,113], as implemented in TeraChem[114,115], to satisfy the RMSD criteria. A dielectric of 10 to mimic the crystal field and defaults of 1.2 × Bondi's van der Waals radii[116] were used to construct the cavity in this case.

Spin-polarized DFT calculations on these nine optimized structures were carried out with the plane-wave periodic boundary condition code Quantum-ESPRESSO[117] v5.1, using the PBE semilocal GGA[11] xc functional. Ultrasoft pseudopotentials[118,119] were employed throughout as obtained from the Quantum-ESPRESSO website with semi-core 3*s* and 3*p* states in the valence for Ti and V (Supporting Information Table S2). Plane wave cutoffs were selected to be 30 Ry for the wavefunction and 300 Ry for the charge density, as in our prior work on similar systems[84]. All complexes were placed with the metal in the center of a 26.5 Å cubic box to ensure sufficient vacuum, and the Martyna-Tuckerman scheme[120] was employed to eliminate periodic image effects. All optimized and re-centered geometries are provided in the Supporting Information .zip file. More than 5 and up to a maximum of 14 unoccupied states (i.e., bands) were included for all calculations. To improve SCF convergence, the mixing factor was reduced to 0.4 from its default value, and the convergence threshold for the SCF energy error was loosened to $9 \times 10^{-6}$ Ry.

All fractional charge calculations were carried out as single point SCF calculations on the optimized geometries of each TM complex. Fractional electron single-point calculations were carried out while varying the net charge such that the formal metal oxidation varied from +4 to +3 for Ti and +5 to +4 for V. All fractional electron calculations were carried out in increments of 0.1 e⁻ by manually altering the band occupations using the "from_input" command in Quantum-ESPRESSO.

Plane-wave eigenstates were transformed to their corresponding real-space Wannier function[121] localized molecular orbitals (MOs) using the pmw.x utility available with the Quantum-ESPRESSO package (Supporting Information Text S2). Five contiguous eigenstates were selected using the reduced state of each complex and their range was specified using the "first_band" and "last_band" keywords (Supporting Information Table S1 and Text S2). To choose the starting index (i.e., the "first_band" value), we noted the band for which the electron would be added or removed (i.e., the majority spin HOMO) as well as four higher energy, majority spin states on the electron configuration. The band numbers depended on the number of atoms as well as the valence states included in the pseudopotentials for each atom type (Supporting Information Tables S1-S2). Density isosurfaces of molecular orbitals were plotted at ±0.002 e/bohr$^3$ using the pp.x utility of Quantum-ESPRESSO and visualized using the VMD visualization program[122] and VESTA[123] v3.3.9.

**3. Results and Discussion.**

**3a. Applying jmDFT on Transition Metal Complexes.**

To determine the suitability of jmDFT as a low-cost method for eliminating both EDE and SCE on molecular systems, we curated a test set of mononuclear octahedral transition metal TM complexes (see Computational Details). We start by using jmDFT in its standard form with atomic projectors and test its suitability for reducing both EDE and SCE for nine representative Ti(III)- and V(IV)-based TM complexes (Figure 1 and Supporting Information Text S1, Table S1 and see Computational Details).

We quantify the EDE as the maximum negative energetic deviation ($E^{dev.}$) from linearity that occurs at the midpoint of the $d^0$ to $d^1$ FCL, as in prior work[84] (Figure 2 and Supporting Information Figure S1). Since constant average curvature is common, this approximation holds well in most cases. We can estimate the maximum deviation using the frontier eigenvalues of the $d^0$ (i.e., $N$–electron) and $d^1$ (i.e., $N+1$–electron) systems[18,26,84]:

$$E^{dev.} = \frac{\varepsilon_N^{LUMO} - \varepsilon_{N+1}^{HOMO}}{8}, \qquad (6)$$

Since a +$U$ correction is a key component in the jmDFT expressions, we expect jmDFT to reduce the EDE for our Ti and V complexes as well. Indeed, using the non-empirical-jmDFT $U$ and $J$ on the metal $d$ atomic projections always leads to a smaller magnitude of the $E^{dev.}$ (< 0.5 eV) relative to the 0.5-0.7 eV EDE obtained with semi-local DFT (i.e., PBE[11], Figure 2 and Supporting Information Tables S3–S4). Still, the EDE reduction remains inefficient (< 50% for most Ti and V complexes) with standard jmDFT (Figure 2 and Supporting Information Table S3). This can expected due to the similarity with the previously observed effect of standard atomic projection DFT+U[84,89] leading to inefficiencies in EDE reduction at non-empirical $U$ values.

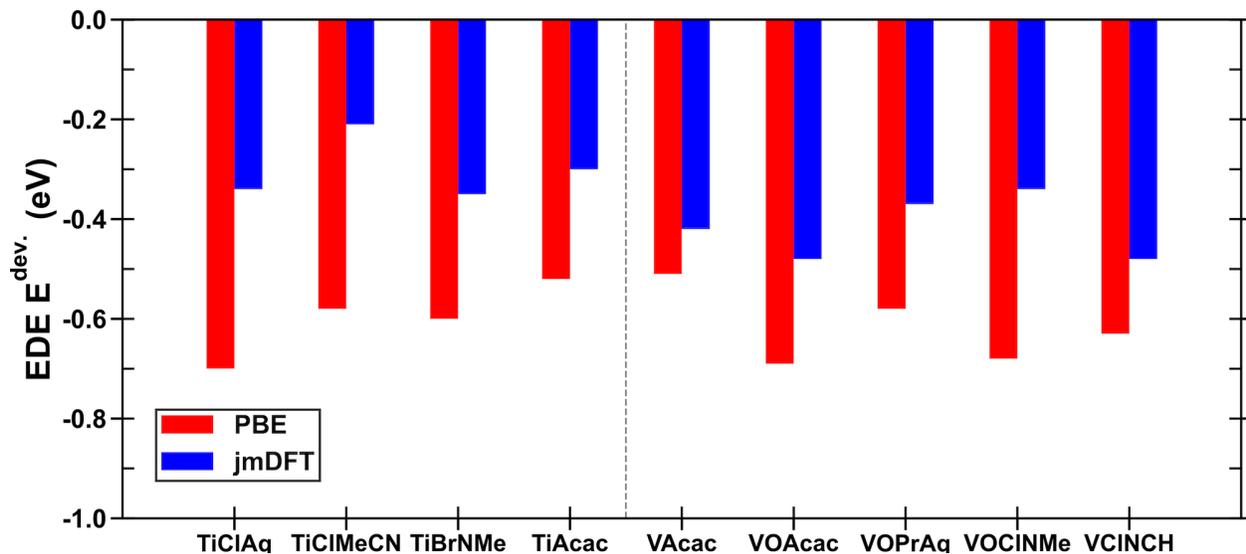

**Figure 2.** Energetic delocalization error (EDE) quantified as the maximum deviation, $E^{dev.}$, from linearity (negative values in eV) occurring at the midpoint of the $d^0$ to $d^1$ FCL for all Ti and V complexes. $E^{dev.}$ computed using the PBE functional is indicated in red bars while $E^{dev.}$ computed with the standard jmDFT correction using non-empirical $U$ and $J$ on the metal $d$ atomic orbitals is indicated in blue bars. All complexes are annotated using the nomenclature introduced earlier, where all Ti complexes are to the left of the vertical grey dashed line and all V complexes are to its right.

Despite its success in reducing EDE, a +U correction is known to increase the SCE in all cases because it penalizes all fractional occupations, even the splitting of an electron between degenerated spin up and down orbitals[85]. The jmDFT approach avoids this by introducing +$J$ and +$J'$ terms that instead favor fractionally occupied spins. We analyze the influence of jmDFT with standard atomic projections on the SCE along the fractional spin line (FSL) of the Ti and V complexes and compare it with the SCE observed with DFT+U alone (Figure 3 and Supporting Information Figure S1 and Tables S4–S5). We quantify the SCE as the maximum positive $E^{dev.}$ from exact energetic invariance that occurs at the midpoint along the FSL (Figure 3). Assuming a constant average curvature along the FSL, we can estimate this value using the frontier eigenvalues of the $d^1$ system alone[26,86]:

$$E^{dev.} = \frac{\varepsilon_{N+1}^{LUMO} - \varepsilon_{N+1}^{HOMO}}{4}. \tag{7}$$

For PBE, the SCE is positive, ranging from 0.2 eV (i.e., for Ti(acac)$_3$) to 0.5 eV (i.e., for VO(acac)$_2$), with comparable values for most complexes (Figure 3 and Supporting Information Table S5). The standard DFT+U increases this SCE further for all complexes, to as large as 1.6 eV (i.e., for VO(acac)$_2$), although the increase is smaller (e.g., 0.4 eV for Ti(acac)$_3$) for other complexes (Supporting Information Table S5). In contrast, applying jmDFT with non-empirical $U$ and $J$ values and standard atomic projectors reduces SCE across most complexes (Figure 3 and Supporting Information Tables S4–S5). The largest SCE reduction (74%) is observed for a V complex (i.e., VOCl$_2$(NH(CH$_3$)$_2$)$_2$), while for Ti complexes, the SCE reduction is only up to 40% (i.e., TiBr$_3$(N(CH$_3$)$_3$)$_2$). These are not element-dependent trends since inefficient SCE reduction was also observed in the case of a V complex (i.e., VCl$_4$(NCH)$_2$) and at its extreme, standard jmDFT showed no effect on Ti(acac)$_3$ (Figure 3 and Supporting Information Table S5).

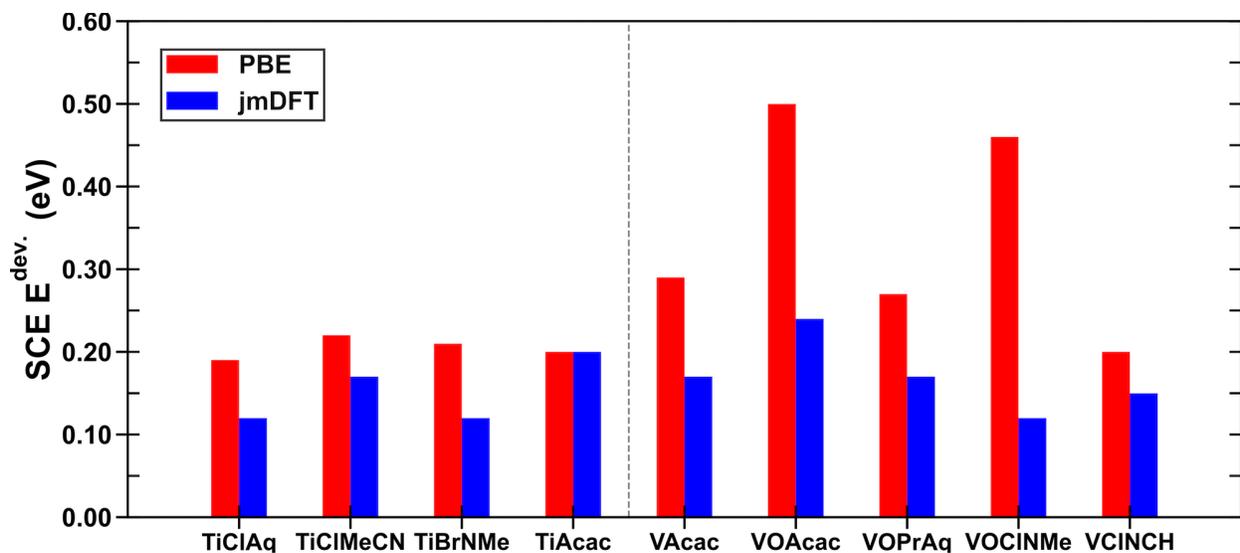

**Figure 3.** Static correlation error (SCE) quantified as the maximum deviation, $E^{dev.}$, from energetic invariance (positive values in eV) occurring at the FSL midpoint for all Ti and V complexes. $E^{dev.}$ computed using the PBE functional is indicated in red bars while $E^{dev.}$ computed with the standard jmDFT correction using non-empirical $U$ and $J$ on the metal $d$ atomic orbitals is indicated in blue bars. All complexes are annotated using the nomenclature introduced earlier, where all Ti complexes are to the left of the vertical grey dashed line and all V complexes are to its right.

Thus, jmDFT holds promise for being one of the few methods that reduces EDE without increasing SCE or computational cost, but we have yet to demonstrate that this elimination occurs at the non-empirical parameter values for TM complexes to the same extent we demonstrated in atoms[86]. In previous work[89], we observed that inefficiencies in DFT+U EDE elimination at non-empirical parameters could be traced to hybridization in molecular complexes in a manner that made atomic projectors unsuitable for the error corrections in the DFT+U functional form. We next investigate if apparent inefficiencies in jmDFT SCE elimination are similarly derived from using atomic projectors.

### 3b. Applying Molecular DFT+J to Transition Metal Complexes.

In previous work[84], we attributed the inefficiencies in DFT+U EDE elimination of molecular complexes to the use of metal-centered atomic orbital projectors (AOPs) within DFT+U. Strong metal-ligand hybridization in the frontier orbital where the fractional electron is added (i.e., along the FCL) made AOPs inefficient for reducing delocalization error[84]. To determine whether inefficiencies in jmDFT SCE elimination also arise due to the use of metal-centered AOPs, we analyze the hybridization in the frontier orbital along the FSL (Figure 4 and Supporting Information Tables S6–S14). For a representative Ti complex, Ti(acac)$_3$, the SCE reduction using AOP-based non-empirical jmDFT was negligible relative to the PBE GGA value (see Sec. 3a, Supporting Information Table S5). We indeed find that the contribution from Ti(3$d$) AOs to the HOMO at the FSL midpoint is less than 50% (Figure 4). The C(2$p$) and O(2$p$) AOs on the ligands contribute comparably to this HOMO, suggesting that metal-centered AOPs alone will be insufficient in describing the fractional spin change along the FSL. Similarly, V complexes (i.e., VCl$_4$(NCH)$_2$) also exhibit strong metal-ligand hybridization with their ligand AOs making a significant (i.e., 32%) contribution to the FSL HOMO (Supporting Information Table S14). This highlights that the use of metal local AOPs alone will limit the SCE reduction efficiency along their FSLs.

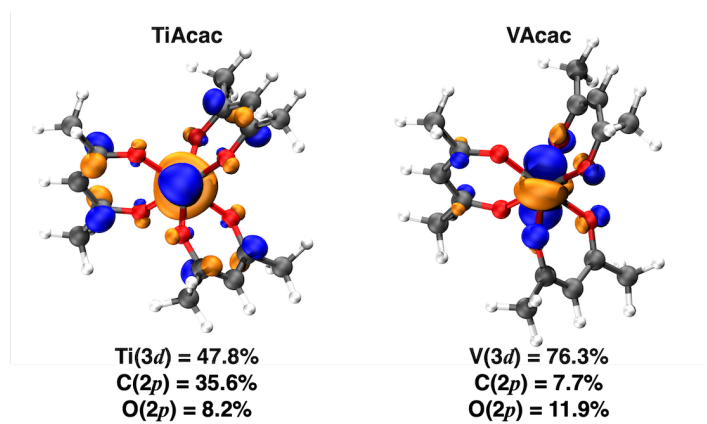

TiAcac
Ti(3$d$) = 47.8%
C(2$p$) = 35.6%
O(2$p$) = 8.2%

VAcac
V(3$d$) = 76.3%
C(2$p$) = 7.7%
O(2$p$) = 11.9%

**Figure 4.** Density isosurface (isovalue = 0.003 e–/bohr$^3$) for the highest occupied molecular orbital (HOMO) of the ($n_\alpha$, $n_\beta$) = ($\frac{1}{2}$, $\frac{1}{2}$) system of Ti(III)(acac)$_3$ (left) and V(IV)(acac)$_3$. Blue denotes the positive wavefunction phase and orange denotes the negative wavefunction phase. Carbon, oxygen and hydrogen atoms are shown using grey, red and white spheres respectively. Atomic orbital (AO) contributions to the HOMO are annotated for Ti(3$d$), V(3$d$), C(2$p$) and O(2p) AOs, where contributions are summed over all atoms of the same type and all angular momenta for each AO.

Based on previous work[84,89], the presence of metal-ligand hybridization suggests that the use of molecular orbitals as projectors (MOPs) may lead to higher SCE reduction efficiency. We use the real-space representations of five frontier MOs from the $d^1$ configuration for each complex as our MOPs (Supporting Information Text S2 and Tables S15–S23). For constructing these MOPs, we select the majority spin, α, MO containing the single unpaired electron in the $d^1$ system as well as the next four higher energy α MOs and the corresponding β MOs from the minority spin (Supporting Information Tables S15–S23).

We then quantified the extent to which these MOPs reduce SCE in strongly hybridized complexes (Figure 5). Since +J correction with a negative J value is used to reduce SCE within jmDFT[85,86], we first investigate SCE reduction only using the +J correction[68,124]:

$$E^{\mathrm{DFT}+J} = E^{\mathrm{DFT}} + \frac{1}{2}\sum_{\sigma} J[\mathrm{Tr}(\mathbf{n}_{\sigma}\mathbf{n}_{-\sigma})]. \tag{8}$$

To a first-order approximation, the PBE fractionality Tr[$\mathbf{n}_\alpha\mathbf{n}_\beta$] determines the efficiency of SCE reduction for a given projector choice. If idealized occupations are obtained, the Tr[$\mathbf{n}_\alpha\mathbf{n}_\beta$] term should be parabolic with a maximum value of 0.25 e$^{-2}$ when the electron is split between α and β orbitals (Figure 5). At odds with this ideal behavior, the Ti(acac)$_3$ AOP fractionalities are nearly constant along the FSL (Figure 5). Correspondingly, the AOP fractionality deviations (i.e., the difference between the midpoint and endpoint) are significantly smaller (< 0.10 e$^{-2}$) than the atomic limit maximum. In contrast, MOP fractionality variations are indeed parabolic along the FSL with a large fractionality deviation (0.23 e$^{-2}$) close to the atomic limit. This suggests the MOPs should be efficient for reducing SCE in Ti(acac)$_3$.

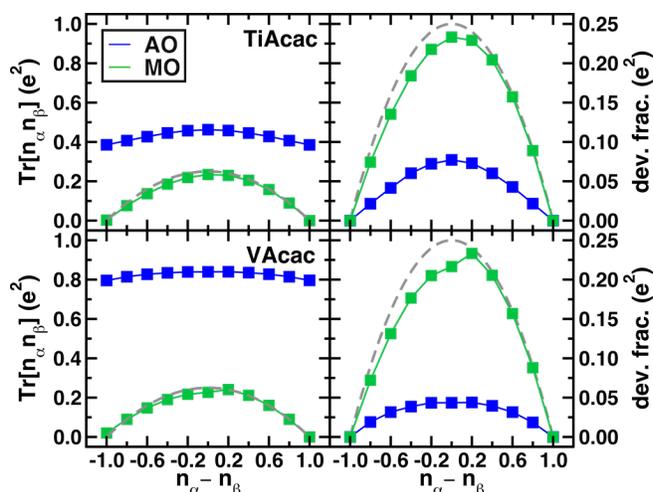

**Figure 5.** PBE GGA fractionalities (i.e., Tr[$\mathbf{n}_\alpha \mathbf{n}_\beta$] in e$^{-2}$, left) and deviations from linear admixture (dev. frac. in e$^{-2}$, right) for TiAcac (top, Ti(acac)$_3$) and VAcac (bottom, V(acac)$_3$), where 'acac' corresponds to the acetylacetonate ligand, along the fractional spin line. Fractionalities and the deviations were obtained using both atomic orbital (AO) projectors (blue squares) and molecular orbital (MO) projectors obtained from the state with formal $d^1$ configuration (green squares). Fractionalities and fractionality deviations at the ideal atomic limit are indicated using grey dashed lines.

To quantify differences in efficiency, we compute the linearized sensitivity in the variation of the SCE with increasingly negative $J$ values, $S_J$(SCE) for both AOPs and MOPs (Figure 6). We indeed observe that MOPs have much higher $S_J$(SCE) and thus, reduce the SCE efficiently for Ti(acac)$_3$ (Figure 6). Other strongly hybridizing complexes (e.g., VCl$_4$(NCH)$_2$) similarly benefit from the use of MOPs, as is clear from higher MOP fractionality deviations and higher MOP-based SCE reduction efficiencies (Supporting Information Table S14 and Figure S2).

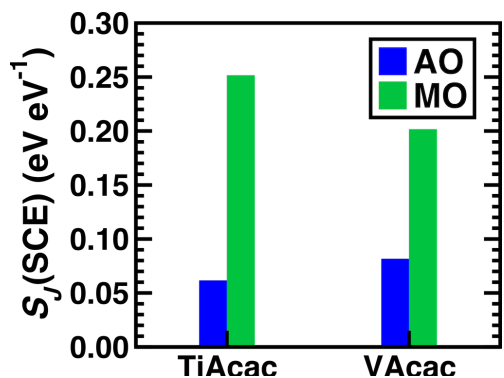

**Figure 6.** Linearized sensitivities to the variation in static correlation error (SCE) with increasingly negative $J$ values at $U = 0$ ($S_J$(SCE), in eV/eV of $J$) evaluated for TiAcac (i.e., Ti(acac)$_3$) and VAcac (i.e., V(acac)$_3$), where 'acac' corresponds to the acetylacetonate ligand using AO projectors (blue bar) and MO projectors (green bar).

When hybridization is not the culprit for low efficiency in SCE or EDE reduction, more subtle changes in occupations between electron configurations had also been observed to be a source for efficiencies smaller than idealized values[84,89] because fractionality deviations were reduced. The V(acac)$_3$, has a significantly higher (> 75%) metal $3d$ AO contribution to the HOMO at its FSL midpoint than Ti(acac)$_3$ (Figure 4 and Supporting Information Table S10). However, we observe that its maximum AOP fractionality deviation (0.04 e$^{-2}$) is not only lower than the atomic limit value but lower than that for Ti(acac)$_3$ (0.08 e$^{-2}$) (Figure 5). In this case, the use of MOPs also results in higher fractionality deviations and $S_J$(SCE) values for V(acac)$_3$ (Figure 6). Thus, MOPs eliminate the SCE across a range of hybridization regimes that would otherwise limit the suitability of jmDFT with AOPs. We next investigate whether jmDFT with non-empirical $U$ and $J$ terms[85] can robustly eliminate both errors (i.e., the flat plane error) for all nine Ti and V complexes.

**3c. Applying Molecular jmDFT on Transition Metal Complexes**

The MO DFT+U approach has previously been observed to be more efficient at EDE reduction than the AO-based approach for a range of hybridizations in $3d$ transition metal complexes[89]. Taken together with the higher efficiency of the MO DFT+J correction towards SCE reduction, we therefore expect that MO jmDFT will be the ideal framework to simultaneously reduce both EDE and SCE. Still, to remain fully non-empirical, we also require that MO-based error reduction efficiencies be close to the theoretically optimal values typically achieved only on model systems[84,86]. We thus first investigate the MO-based EDE and SCE reduction efficiencies for all complexes relative to their idealized limit values[84,86] before analyzing how well this translates to an elimination of both errors within MO jmDFT.

For all nine Ti and V complexes, we computed both the SCE reduction efficiencies using $S_J$(SCE), as defined earlier, and the EDE reduction efficiency using $S_U$(EDE) evaluated with increasingly positive $U$ values (Figure 7). The optimal efficiencies for EDE and SCE reduction on model systems are 0.125 eV per eV of $U$[84] for $S_U$(EDE) and 0.25 eV per eV of $J$[86] for $S_J$(SCE). As expected, the MO approach reduces errors more efficiently relative to the AO approach for both EDE and SCE across all Ti and V complexes, achieving close to the idealized limit (Figure 7). The MO approach achieves universally higher $S_J$(SCE) values, with all at least 80% of the idealized limit (i.e., 0.20 eV/eV or above). The MO-based $S_U$(EDE) varies more widely, with most Ti complexes having values closer to the idealized limit value of 0.125 eV/eV. Nevertheless, some V complexes have lower MO $S_U$(EDE) values due to shifts in frontier orbital ordering at moderate $U$ values (e.g., V(acac)$_3$, Supporting Information Figure S3). Nevertheless, the broadly good sensitivities obtained with the MO approach suggest that MO jmDFT should largely reduce both errors simultaneously at the non-empirical jmDFT limit.

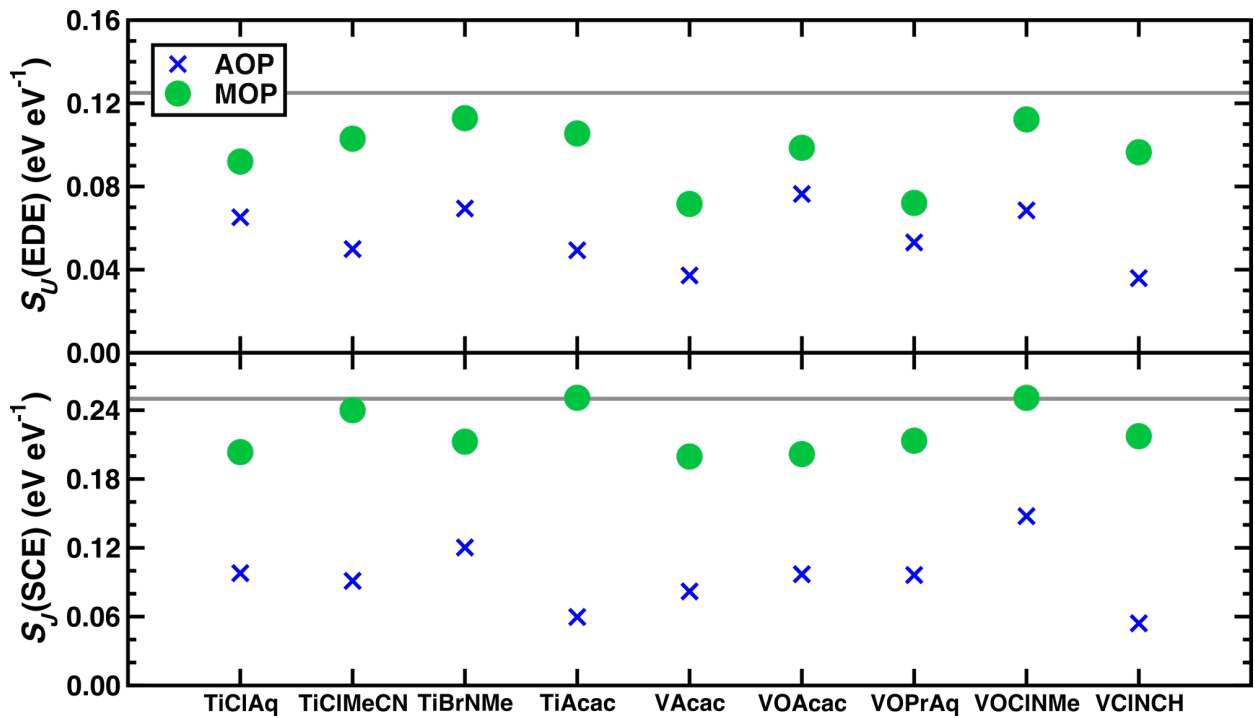

**Figure 7.** Linearized sensitivities to the variation in energetic delocalization error (EDE, top) and in static correlation error (SCE, bottom) with increasingly positive $U$ values at $J = 0$ eV ($S_U$(EDE), in eV per eV of $U$) and increasingly negative $J$ values at $U = 0$ eV ($S_J$(SCE), in eV per eV of $J$) respectively for all Ti and V complexes under study. Sensitivities were evaluated using AO projectors (blue crosses) and MO projectors (green circles). Theoretical maximum values of each sensitivity at the atomic limit are indicated using a solid gray line.

Finally, we computed EDE and SCE at the non-empirical jmDFT limit using the MOP approach (Figure 8). We observe that both errors are largely reduced across most Ti and V complexes. Consistent with the sensitivity analysis, EDEs were virtually eliminated for the Ti complexes, with a larger than 80% reduction (Figures 7 and 8). Similarly, EDE reduction for the V complexes was slightly less effective (e.g., VAcac) due to slightly lower MO $S_U$(EDE) values, but these reductions are still larger than obtained from the AOP approach (Supporting Information Table S24). The universally higher $S_J$(SCE) values with the MOP approach indeed translate to a simultaneously large reduction in the SCE. For example, the MOP approach nearly eliminates the SCE for Ti(acac)$_3$ and VOCl$_2$(NH(CH$_3$)$_2$)$_2$ (Figure 8). For V complexes, SCE reduction with the MOP jmDFT correction is far more consistent than it was with AOPs, generally reducing the SCE by greater than 70% (Supporting Information Table S24). This good performance includes cases where the AOP approach was already observed to be relatively efficient as in $VOCl_2(NH(CH_3)_2)_2$. Similarly, for Ti complexes, SCEs are nearly eliminated using the MOP jmDFT approach except for the case of $TiBr_3(N(CH_3)_3)_2$ due to a change in frontier orbital ordering with $J$ values close to the non-empirical values (Supporting Information Figure S4). Overall, we observe that the MOP-based jmDFT correction offers an efficient way for recovering exact properties in nearly all TM complexes within a non-empirical framework at the low cost of semi-local DFT.

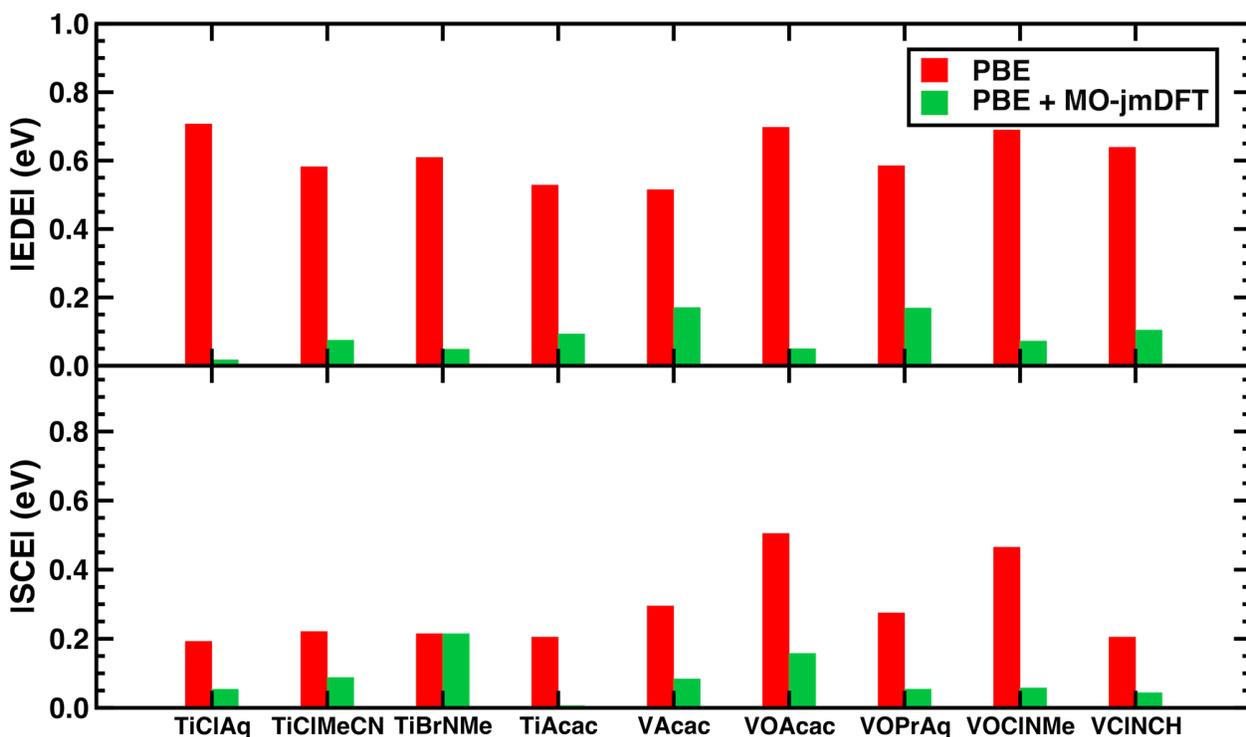

**Figure 8.** (top) Energetic delocalization error (EDE, top) quantified as the absolute maximum deviation from linearity (in eV) at the midpoint of the $d^0$ to $d^1$ FCL for all Ti and V complexes. (bottom) Static correlation error (SCE, bottom) quantified as the absolute maximum deviation from energetic invariance (in eV) at the FSL midpoint for all Ti and V complexes. All errors are computed using the PBE functional (red bars) and with the MO-projector based jmDFT correction using non-empirical $U$ and $J$ values (green bars).

**4. Conclusions.**

Here, we demonstrate an approach to recovering exact conditions at semi-local DFT cost with the jmDFT method. By using frontier orbital energies as input coefficients in the jmDFT +U and +J expressions, the method has the promise of being fully non-empirical. While the approach in its non-empirical form was previously introduced and demonstrated on atoms and small diatomics, we have now extended it to show the method works equally well at recovering exact conditions in transition metal complexes. First, we curated a set of nine representative Ti(III) and V(IV) $d^1$ transition metal complexes. We then analyzed the errors from PBE of these complexes along the fractional spin line of the $d^1$ complex as well as the fractional charge line between $d^0$ and $d^1$ electron configurations. We rationalized inefficiencies observed for standard (i.e., AOP) jmDFT along the fractional charge and spin lines by noting hybridization in the relevant frontier orbitals. This led to lower than expected fractional occupations detected from the projections on the FSL, an observation we had previously only made along the FCL. To overcome these limitations, we introduced a procedure for computing an MOP basis for jmDFT. We demonstrated this single set of $d^1$ MOPs to be suitable for nearly eliminating all energetic delocalization error and static correlation error along the FCL and FSL, respectively. In all cases, the MOP jmDFT outperformed AOP jmDFT. Furthermore, unlike DFT+U, jmDFT never improved energetic delocalization error at the cost of worsening static correlation error. We expect this non-empirical approach and systematic construction of jmDFT MOPs to be broadly useful in high-throughput screening in transition metal chemistry.

ASSOCIATED CONTENT

**Supporting Information**. Extraction of transition metal complexes with $d^1$ configuration from literature; CSD refcodes and chemical formulae of the final 9 complexes; List of pseudopotentials used; Construction of real-space molecular orbitals; A representative exact flat-plane construction; Energetic delocalization error (EDE) at the PBE and atomic jmDFT levels; Non-empirical values of U and J for all nine Ti and V complexes; Static correlation error (SCE) at the PBE, PBE+U and atomic jmDFT levels; Atomic orbital character of frontier spin-up MOs for the FSL of TiClAq; Atomic orbital character of frontier spin-up MOs for the FSL of TiClMeCN; Atomic orbital character of frontier spin-up MOs for the FSL of TiBrNMe; Atomic orbital character of frontier spin-up MOs for the FSL of TiAcac; Atomic orbital character of frontier spin-up MOs for the FSL of VAcac; Atomic orbital character of frontier spin-up MOs for the FSL of VOAcac; Atomic orbital character of frontier spin-up MOs for the FSL of VOPrAq; Atomic orbital character of frontier spin-up MOs for the FSL of VOClNMe; Atomic orbital character of frontier spin-up MOs for the FSL of VClNCH; Atomic orbital character of the projector MOs of TiClAq; Atomic orbital character of the projector MOs of TiClMeCN; Atomic orbital character of the projector MOs of TiBrNMe; Atomic orbital character of the projector MOs of TiAcac; Atomic orbital character of the projector MOs of VAcac; Atomic orbital character of the projector MOs of VOAcac; Atomic orbital character of the projector MOs of VOPrAq; Atomic orbital character of the projector MOs of VOClNMe; Atomic orbital character of the projector MOs of VClNCH; FSL fractionality analysis and SCE reduction efficiencies for VClNCH; Frontier eigenvalue variations for VAcac with U using MO projectors; EDE and SCE values for PBE, atomic jmDFT and molecular jmDFT; Frontier eigenvalue variations for TiBrNMe with J using MO projectors (PDF)

Structures of all molecules studied in this work (ZIP)

AUTHOR INFORMATION

**Corresponding Author**

*email: hjkulik@mit.edu phone: 617-253-4584**Notes**

The authors declare no competing financial interest.

ACKNOWLEDGMENT

This work was primarily supported by the U.S. Department of Energy under grant number DE-SC0018096. Workflows used in this work were supported under Office of Naval Research Grant numbers N00014-18-1-2434 and N00014-20-1-2150 (C.D. and M.G.T) and the national science foundation grant number CBET-1846426 (A.N.). This work was also partially supported by a National Science Foundation Graduate Research Fellowship under Grant #1122374 (to A.N.). H.J.K. holds a Career Award at the Scientific Interface from the Burroughs Wellcome Fund, an AAAS Marion Milligan Mason Award, and an Alfred P. Sloan award in Chemistry, which supported this work. This work made use of the Extreme Science and Engineering Discovery Environment (XSEDE), which is supported by National Science Foundation grant number ACI-1548562. The authors thank Adam H. Steeves for providing a critical reading of the manuscript.

# Supporting Information for

*Molecular orbital projectors in non-empirical jmDFT recover exact conditions in transition metal chemistry*

Akash Bajaj[1,2], Chenru Duan[1,3], Aditya Nandy[1,3], Michael G. Taylor[1], and Heather J. Kulik[1,*]

[1]Department of Chemical Engineering, Massachusetts Institute of Technology, Cambridge, MA 02139

[2]Department of Materials Science and Engineering, Massachusetts Institute of Technology, Cambridge, MA 02139

[3]Department of Chemistry, Massachusetts Institute of Technology, Cambridge, MA 02139

**Contents**

**Text S1.** Extraction of transition metal complexes with $d^1$ configuration from literature.

We searched for mononuclear transition metal (TM) complexes that were expected to have a formal $d^1$ electron configuration using the Cambridge Structural Database[1] (CSD version 5.41, Nov. 2019 + 3 data updates to August 2020). We performed an initial search for mononuclear TM complexes with Ti or V as the metal centers with (i) a maximum of 50 atoms in total, (ii) less than half of which were allowed to be non-hydrogen heavy atoms, and with (iii) user-labeled oxidation states of +3 for Ti and +4 for V (n = 74 for Ti, n = 303 for V). We excluded all other $3d$ transition metals since their high-valent $d^1$ configuration complexes were rare (n = 30 for Sc, Cr-Zn). For the remaining complexes, we calculated their individual ligand charges based on the user-uploaded connectivity information and the octet rule. We next selected neutral (i.e., zero net charge) Ti/V complexes and verified the consistency between their total charge, ligand charges and the metal oxidation state i.e., the sum over the metal oxidation state and each individual ligand charge should be the same as the total complex charge (n = 126). From these, 50 structures were manually chosen, to cover representative examples for different metal-coordinating atom environments (i.e., N, O, Cl and Br). When there were duplicates, we selected the structure with the lowest CSD R-factor. Additionally, we incorporated the neutral V(acac)$_3$ complex (refcode: VAACAC23) for comparison with its extracted Ti analogue (refcode: XAQWIY).

**Table S1.** Chemical formulae and Cambridge Structural Database (CSD) refcodes for the nine representative Ti/V mononuclear transition metal complexes selected in the final dataset that exhibited high geometric fidelity (i.e., root mean squared deviation < 0.4 Å) relative to the CSD structure after geometry optimization. Unique ligand atoms directly coordinating to the metal are indicated separately. In the listed chemical formulae, 'acac' corresponds to the acetylacetonate ligand while 'prz' corresponds to the pyrazinoic acid (deprotonated) ligand. Values specified for the "first_band" and "last_band" keywords for generating real-space Wannier function molecular orbitals are indicated.

| CSD refcode | Chemical Formula | Metal Coordinating Atoms | First Band Index | Last Band Index |
|---|---|---|---|---|
| ACACVO14 | V(=O)(acac)$_2$ | O | 49 | 53 |
| CFMNTV | VCl$_4$(NCH)$_2$ | N, Cl | 31 | 35 |
| DUDKUI11 | TiCl$_3$(CH$_3$CN)$_3$ | N, Cl | 41 | 45 |
| IHOQIG | V(=O)Cl$_2$(NH(CH$_3$)$_2$)$_2$ | N, O, Cl | 37 | 41 |
| NAVMAY | TiCl$_3$(H$_2$O)$_3$ | O, Cl | 29 | 33 |
| TMATBI | TiBr$_3$(N(CH$_3$)$_3$)$_2$ | N, Br | 43 | 47 |
| VAACAC23 | V(acac)$_3$ | O | 65 | 69 |
| VOKRAQ | V(=O)(prz)$_2$(H$_2$O) | N, O | 59 | 63 |
| XAQWIY | Ti(acac)$_3$ | O | 65 | 69 |

**Table S2.** List of ultrasoft pseudopotentials employed in this work with filenames identical to those obtained from the Quantum-ESPRESSO website. The valence electrons that are explicitly modeled are indicated with any semi-core states underlined where applicable.

| Element | Pseudopotential Name | Valence States |
|---------|---------------------|----------------|
| H | H.pbe-rrkjus.UPF | $1s$ |
| C | C.pbe-rrkjus.UPF | $2s, 2p$ |
| N | N.pbe-rrkjus.UPF | $2s, 2p$ |
| O | O.pbe-rrkjus.UPF | $2s, 2p$ |
| Cl | Cl.pbe-n-rrkjus_psl.0.1.UPF | $3s, 3p$ |
| Ti | Ti.pbe-sp-van_ak.UPF | $\underline{3s}, \underline{3p}, 3d, 4s$ |
| V | V.pbe-sp-van.UPF | $\underline{3s}, \underline{3p}, 3d, 4s, 4p$ |
| Br | Br.pbe-n-rrkjus_psl.0.2.UPF | $4s, 4p$ |

**Text S2.** Construction of real-space molecular orbitals (MOs).

The transformation of plane-wave molecular states $|\psi_{\mathbf{k},v}\rangle$, where $\mathbf{k}$ denotes the *k*-point and $v$ denotes the band index, to their real-space molecular orbital (MO) counterparts was carried out using the **pmw.x** utility available with the Quantum-ESPRESSO package[2]. From the total number of available eigenstates, we select a specific set by providing a contiguous range for the band index i.e., $v_i$ to $v_f$. The procedure followed by **pmw.x** after providing it with this selection is as follows:

The selected plane-wave states $\{|\psi_{\mathbf{k},v}\rangle\}$ are first projected on the atomic orbitals (AOs) $\{|\phi^I_{\lambda'}\rangle\}$ localized on a Hubbard site *I*:

$$|\tilde{\psi}^I_{\mathbf{k},\lambda'}\rangle = \sum_{v=v_i}^{v_f} |\psi_{\mathbf{k},v}\rangle \langle \psi_{\mathbf{k},v} | \phi^I_{\lambda'} \rangle, \qquad (1)$$

to obtain a set of transformed states $\{|\tilde{\psi}^I_{\mathbf{k},\lambda'}\rangle\}$. For our Ti/V transition metal complexes, the AOs were the five valence $3d$ atomic orbitals associated with the Ti/V metal center that were obtained from the all-electron calculation carried out for pseudopotential generation.

The transformed states are then orthonormalized using the Löwdin symmetric orthonormalization procedure:

$$|\bar{\psi}^I_{\mathbf{k},\lambda}\rangle = \sum_{\lambda'} |\tilde{\psi}^I_{\mathbf{k},\lambda'}\rangle (S_{\mathbf{k}}^{-\frac{1}{2}})_{\lambda'\lambda}, \qquad (2)$$

where $S_{\mathbf{k}}$, the overlap matrix, is defined as $(S_{\mathbf{k}})_{\lambda'\lambda} = \langle \tilde{\psi}^I_{\mathbf{k},\lambda'} | \tilde{\psi}^I_{\mathbf{k},\lambda} \rangle$. The normalized states $\{|\bar{\psi}^I_{\mathbf{k},\lambda}\rangle\}$ are then Fourier transformed from their **k**-space representation to their real-space representation to obtain the desired Wannier functions for the corresponding Hubbard site *I*:

$$|\mathbf{R}^I_\lambda\rangle = \frac{V}{(2\pi)^3} \int_{BZ} d\mathbf{k}\, e^{-i\mathbf{k}\cdot\mathbf{R}} |\bar{\psi}^I_{\mathbf{k},\lambda}\rangle, \qquad (3)$$

where *V* represents the volume of the primitive unit cell in real space.

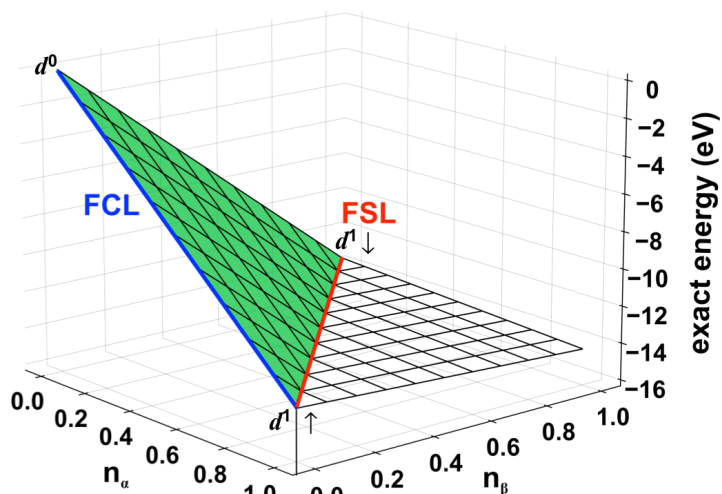

**Figure S1.** Representative flat plane for an exact energy functional as a function of the spin-up electron count ($n_\alpha$) and spin-down electron count ($n_\beta$). The green shaded region denotes the flat-plane region studied in this work for $d^1$ (i.e., Ti and V) transition metal complexes. The formal electron configuration of the metal center in these complexes at each integer-electron endpoint is annotated i.e., $d^0$ for the $N-1$ electron singlet states where $(n_\alpha, n_\beta) = (0, 0)$ and $d^1$ for the two $N$ electron doublet states where $(n_\alpha, n_\beta) = (1, 0)$ and $(0, 1)$. The fractional charge line (FCL) connecting the $d^0$ and $d^1$ systems used for quantifying the energetic delocalization error (EDE) and the fractional spin line (FSL, red) connecting the two spin configurations of the $d^1$ system used for quantifying the static correlation error (SCE) are highlighted in blue and red respectively.

**Table S3.** Energetic delocalization error computed using the maximum deviation, $E^{dev.}$, from linearity (negative values, in eV) occurring at the midpoint of the fractional charge line (FCL) where an electron gets added from the $d^0$ to the $d^1$ configuration. $E^{dev.}$ is computed at the semi-local DFT (i.e., PBE[3] GGA) level and with standard (i.e., atomic projections) non-empirical-jmDFT. The chemical formula of each complex as well as the nomenclature assigned to it throughout this work are both indicated. In the listed chemical formulae, 'acac' corresponds to the acetylacetonate ligand while 'prz' corresponds to the pyrazinoate ligand.

| Nomenclature | Chemical Formula | $E^{dev.}$ along FCL (eV) | |
|---|---|---|---|
| | | PBE | Atomic jmDFT |
| TiClAq | $TiCl_3(H_2O)_3$ | -0.70 | -0.34 |
| TiClMeCN | $TiCl_3(CH_3CN)_3$ | -0.58 | -0.21 |
| TiBrNMe | $TiBr_3(N(CH_3)_3)_2$ | -0.60 | -0.35 |
| TiAcac | $Ti(acac)_3$ | -0.52 | -0.30 |
| VAcac | $V(acac)_3$ | -0.51 | -0.42 |
| VOAcac | $VO(acac)_2$ | -0.69 | -0.48 |
| VOPrAq | $VO(prz)_2(H_2O)$ | -0.58 | -0.37 |
| VOClNMe | $VOCl_2(NH(CH_3)_2)_2$ | -0.68 | -0.34 |
| VClNCH | $VCl_4(NCH)_2$ | -0.63 | -0.48 |

**Table S4.** Non-empirical values of $U$ and $J$ (in eV) evaluated for all nine Ti and V complexes with standard jmDFT[4]. The chemical formula of each complex as well as the nomenclature assigned to it throughout this work are both indicated. In the listed chemical formulae, 'acac' corresponds to the acetylacetonate ligand while 'prz' corresponds to the pyrazinoate ligand.

| Nomenclature | Chemical Formula | Non-empirical $U$ (eV) | Non-empirical $J$ (eV) |
|---|---|---|---|
| TiClAq | $TiCl_3(H_2O)_3$ | 5.75 | -6.50 |
| TiClMeCN | $TiCl_3(CH_3CN)_3$ | 4.64 | -5.50 |
| TiBrNMe | $TiBr_3(N(CH_3)_3)_2$ | 5.00 | -5.84 |
| TiAcac | $Ti(acac)_3$ | 4.14 | -4.85 |
| VAcac | $V(acac)_3$ | 4.31 | -5.48 |
| VOAcac | $VO(acac)_2$ | 5.77 | -6.72 |
| VOPrAq | $VO(prz)_2(H_2O)$ | 4.87 | -5.93 |
| VOClNMe | $VOCl_2(NH(CH_3)_2)_2$ | 5.64 | -7.47 |
| VClNCH | $VCl_4(NCH)_2$ | 5.12 | -5.94 |

**Table S5.** Static correlation error computed using the maximum deviation, $E^{dev.}$, from energetic invariance (positive values, in eV) occurring at the midpoint of the fractional spin line (FSL) connecting the α-spin and the β-spin configurations of the $d^1$ system. $E^{dev.}$ is computed at the semi-local DFT (i.e., PBE[3] GGA) level, with the standard (i.e., atomic projections) DFT+U approach[5-8] using the non-empirical curvature[9-10] $U$ and with standard (i.e., atomic projections) non-empirical-jmDFT. The chemical formula of each complex and the nomenclature assigned to it throughout this work are both indicated. In the listed chemical formulae, 'acac' corresponds to the acetylacetonate ligand while 'prz' corresponds to the pyrazinoate ligand.

| Nomenclature | Chemical Formula | $E^{dev.}$ along FSL (eV) | | |
|---|---|---|---|---|
| | | PBE | PBE+U | Atomic jmDFT |
| TiClAq | $TiCl_3(H_2O)_3$ | 0.19 | 0.78 | 0.12 |
| TiClMeCN | $TiCl_3(CH_3CN)_3$ | 0.22 | 0.66 | 0.17 |
| TiBrNMe | $TiBr_3(N(CH_3)_3)_2$ | 0.21 | 0.94 | 0.12 |
| TiAcac | $Ti(acac)_3$ | 0.20 | 0.54 | 0.20 |
| VAcac | $V(acac)_3$ | 0.29 | 0.70 | 0.17 |
| VOAcac | $VO(acac)_2$ | 0.50 | 1.62 | 0.24 |
| VOPrAq | $VO(prz)_2(H_2O)$ | 0.27 | 1.30 | 0.17 |
| VOClNMe | $VOCl_2(NH(CH_3)_2)_2$ | 0.46 | 1.13 | 0.12 |
| VClNCH | $VCl_4(NCH)_2$ | 0.20 | 0.69 | 0.15 |

**Table S6.** Atomic orbital (AO) contributions to the frontier orbital on the FSL of TiClAq (TiCl$_3$(H$_2$O)$_3$). AO contributions are indicated for the α-spin HOMO of the $(n_\alpha, n_\beta) = (\frac{1}{2}, \frac{1}{2})$ singlet state having a formal $d^1$ configuration at the metal. Contributions are summed by ligand element type but distinguished by principal quantum number and subshell. A sum less than 100% indicates a lack of suitable projection of the extended state onto the all-electron atomic states from the pseudopotentials.

| Atomic Orbitals | α:HOMO $(n_\alpha, n_\beta) = (\frac{1}{2}, \frac{1}{2})$ |
|---|---|
| Ti(3s) | 0 |
| Ti(3p) | 0 |
| Ti(3d) | 78.3% |
| Ti(4s) | 0.6% |
| Cl(3s) | 0 |
| Cl(3p) | 7.7% |
| O(2s) | 2.3% |
| O(2p) | 0.3% |
| H(1s) | 5.4% |
| Total | 94.6% |

**Table S7.** Atomic orbital (AO) contributions to the frontier orbital on the FSL of TiClMeCN (TiCl$_3$(CH$_3$CN)$_3$). AO contributions are indicated for the α-spin HOMO of the $(n_\alpha, n_\beta) = (\frac{1}{2}, \frac{1}{2})$ singlet state having a formal $d^1$ configuration at the metal. Contributions are summed by ligand element type but distinguished by principal quantum number and subshell. A sum less than 100% indicates a lack of suitable projection of the extended state onto the all-electron atomic states from the pseudopotentials.

| Atomic Orbitals | α:HOMO $(n_\alpha, n_\beta) = (\frac{1}{2}, \frac{1}{2})$ |
|---|---|
| Ti(3s) | 0 |
| Ti(3p) | 0 |
| Ti(3d) | 62.0% |
| Ti(4s) | 0 |
| Cl(3s) | 0 |
| Cl(3p) | 7.3% |
| C(2s) | 0 |
| C(2p) | 19.0% |
| N(2s) | 0 |
| N(2p) | 5.6% |
| H(1s) | 3.1% |
| Total | 97.0% |

**Table S8.** Atomic orbital (AO) contributions to the frontier orbital on the FSL of TiBrNMe (TiBr$_3$(N(CH$_3$)$_3$)$_2$). AO contributions are indicated for the α-spin HOMO of the $(n_\alpha, n_\beta) = (\frac{1}{2}, \frac{1}{2})$ singlet state having a formal $d^1$ configuration at the metal. Contributions are summed by ligand element type but distinguished by principal quantum number and subshell. A sum less than 100% indicates a lack of suitable projection of the extended state onto the all-electron atomic states from the pseudopotentials.

| Atomic Orbitals | α:HOMO $(n_\alpha, n_\beta) = (\frac{1}{2}, \frac{1}{2})$ |
|---|---|
| Ti(3$s$) | 0 |
| Ti(3$p$) | 0 |
| Ti(3$d$) | 87.2% |
| Ti(4$s$) | 0 |
| Br(4$s$) | 0 |
| Br(4$p$) | 7.9% |
| C(2$s$) | 0 |
| C(2$p$) | 0.4% |
| N(2$s$) | 0 |
| N(2$p$) | 0 |
| H(1$s$) | 0.2% |
| Total | 95.7% |

**Table S9.** Atomic orbital (AO) contributions to the frontier orbital on the FSL of TiAcac (Ti(acac)$_3$), where 'acac' corresponds to the acetylacetonate ligand. AO contributions are indicated for the α-spin HOMO of the $(n_\alpha, n_\beta) = (\frac{1}{2}, \frac{1}{2})$ singlet state having a formal $d^1$ configuration at the metal. Contributions are summed by ligand element type but distinguished by principal quantum number and subshell. A sum less than 100% indicates a lack of suitable projection of the extended state onto the all-electron atomic states from the pseudopotentials.

| Atomic Orbitals | α:HOMO $(n_\alpha, n_\beta) = (\frac{1}{2}, \frac{1}{2})$ |
|---|---|
| Ti(3$s$) | 0 |
| Ti(3$p$) | 0 |
| Ti(3$d$) | 47.8% |
| Ti(4$s$) | 0 |
| C(2$s$) | 0 |
| C(2$p$) | 35.6% |
| O(2$s$) | 0 |
| O(2$p$) | 8.2% |
| H(1$s$) | 4.7% |
| Total | 96.3% |

**Table S10.** Atomic orbital (AO) contributions to the frontier orbital on the FSL of VAcac (V(acac)$_3$), where 'acac' corresponds to the acetylacetonate ligand. AO contributions are indicated for the α-spin HOMO of the $(n_\alpha, n_\beta) = (\frac{1}{2}, \frac{1}{2})$ singlet state having a formal $d^1$ configuration at the metal. Contributions are summed by ligand element type but distinguished by principal quantum number and subshell. A sum less than 100% indicates a lack of suitable projection of the extended state onto the all-electron atomic states from the pseudopotentials.

| Atomic Orbitals | α:HOMO $(n_\alpha, n_\beta) = (\frac{1}{2}, \frac{1}{2})$ |
|---|---|
| V(3$s$) | 0 |
| V(3$p$) | 0 |
| V(3$d$) | 76.3% |
| V(4$s$) | 0 |
| V(4$p$) | 0 |
| C(2$s$) | 0 |
| C(2$p$) | 7.7% |
| O(2$s$) | 0.6% |
| O(2$p$) | 11.9% |
| H(1$s$) | 0.8% |
| Total | 97.3% |

**Table S11.** Atomic orbital (AO) contributions to the frontier orbital on the FSL of VOAcac (VO(acac)$_2$), where 'acac' corresponds to the acetylacetonate ligand. AO contributions are indicated for the α-spin HOMO of the $(n_\alpha, n_\beta) = (\frac{1}{2}, \frac{1}{2})$ singlet state having a formal $d^1$ configuration at the metal. Contributions are summed by ligand element type but distinguished by principal quantum number and subshell. A sum less than 100% indicates a lack of suitable projection of the extended state onto the all-electron atomic states from the pseudopotentials.

| Atomic Orbitals | α:HOMO $(n_\alpha, n_\beta) = (\frac{1}{2}, \frac{1}{2})$ |
|---|---|
| V(3$s$) | 0 |
| V(3$p$) | 0 |
| V(3$d$) | 87.8% |
| V(4$s$) | 0.2% |
| V(4$p$) | 0 |
| C(2$s$) | 0.1% |
| C(2$p$) | 1.1% |
| O(2$s$) | 0.6% |
| O(2$p$) | 8.0% |
| H(1$s$) | 0.1% |
| Total | 97.9% |

**Table S12.** Atomic orbital (AO) contributions to the frontier orbital on the FSL of VOPrAq (VO(prz)$_2$(H$_2$O)), where 'prz' corresponds to the pyrazinoate ligand. AO contributions are indicated for the α-spin HOMO of the $(n_\alpha, n_\beta) = (\frac{1}{2}, \frac{1}{2})$ singlet state having a formal $d^1$ configuration at the metal. Contributions are summed by ligand element type but distinguished by principal quantum number and subshell. A sum less than 100% indicates a lack of suitable projection of the extended state onto the all-electron atomic states from the pseudopotentials.

| Atomic Orbitals | α:HOMO $(n_\alpha, n_\beta) = (\frac{1}{2}, \frac{1}{2})$ |
|---|---|
| V(3s) | 0 |
| V(3p) | 0 |
| V(3d) | 70.1% |
| V(4s) | 0 |
| V(4p) | 0.3% |
| C(2s) | 0 |
| C(2p) | 11.7% |
| N(2s) | 0 |
| N(2p) | 9.0% |
| O(2s) | 0.1% |
| O(2p) | 6.8% |
| H(1s) | 0.2% |
| Total | 98.2% |

**Table S13.** Atomic orbital (AO) contributions to the frontier orbital on the FSL of VOClNMe (VOCl$_2$(NH(CH$_3$)$_2$)$_2$). AO contributions are indicated for the α-spin HOMO of the $(n_\alpha, n_\beta) = (\frac{1}{2}, \frac{1}{2})$ singlet state having a formal $d^1$ configuration at the metal. Contributions are summed by ligand element type but distinguished by principal quantum number and subshell. A sum less than 100% indicates a lack of suitable projection of the extended state onto the all-electron atomic states from the pseudopotentials.

| Atomic Orbitals | α:HOMO $(n_\alpha, n_\beta) = (\frac{1}{2}, \frac{1}{2})$ |
|---|---|
| V(3$s$) | 0 |
| V(3$p$) | 0 |
| V(3$d$) | 86.9% |
| V(4$s$) | 0 |
| V(4$p$) | 0 |
| Cl(3$s$) | 0 |
| Cl(3$p$) | 9.7% |
| C(2$s$) | 0 |
| C(2$p$) | 0 |
| N(2$s$) | 0 |
| N(2$p$) | 0 |
| O(2$s$) | 0 |
| O(2$p$) | 0 |
| H(1$s$) | 1.3% |
| Total | 97.9% |

**Table S14.** Atomic orbital (AO) contributions to the frontier orbital on the FSL of VClNCH (VCl$_4$(NCH)$_2$). AO contributions are indicated for the α-spin HOMO of the $(n_\alpha, n_\beta) = (\frac{1}{2}, \frac{1}{2})$ singlet state having a formal $d^1$ configuration at the metal. Contributions are summed by ligand element type but distinguished by principal quantum number and subshell. A sum less than 100% indicates a lack of suitable projection of the extended state onto the all-electron atomic states from the pseudopotentials.

| Atomic Orbitals | α:HOMO $(n_\alpha, n_\beta) = (\frac{1}{2}, \frac{1}{2})$ |
|---|---|
| V(3s) | 0 |
| V(3p) | 0 |
| V(3d) | 68.3% |
| V(4s) | 0 |
| V(4p) | 0 |
| Cl(3s) | 0 |
| Cl(3p) | 24.9% |
| C(2s) | 0 |
| C(2p) | 3.9% |
| N(2s) | 0.4% |
| N(2p) | 0.7% |
| H(1s) | 0 |
| Total | 98.2% |

**Table S15.** Atomic orbital (AO) contributions to the five molecular states (i.e., the α-HOMO and four higher energy MOs) used for constructing real-space molecular orbital (MO) projectors for each spin for TiClAq (TiCl$_3$(H$_2$O)$_3$) in its $(n_\alpha, n_\beta) = (1, 0)$ $d^1$ doublet state along with the β LUMO and four higher energy β MOs. Contributions are summed by ligand element type but distinguished by principal quantum number and subshell. A sum less than 100% indicates a lack of suitable projection of the extended state onto the all-electron atomic states from the pseudopotentials.

| Atomic Orbitals | α | | | | | β | | | | |
|---|---|---|---|---|---|---|---|---|---|---|
| | HOMO | LUMO | LUMO+1 | LUMO+2 | LUMO+3 | LUMO | LUMO+1 | LUMO+2 | LUMO+3 | LUMO+4 |
| Ti(3s) | 0 | 0 | 0 | 0 | 0 | 0 | 0 | 0 | 0 | 0 |
| Ti(3p) | 0 | 0 | 0 | 0 | 0 | 0 | 0 | 0 | 0 | 0 |
| Ti(3d) | 82.4% | 78.8% | 84.8% | 5.8% | 34.5% | 45.3% | 85.6% | 39.2% | 76.6% | 20.5% |
| Ti(4s) | 0.1% | 0.3% | 0.1% | 1.3% | 0 | 1.2% | 0 | 0.5% | 0 | 0 |
| Cl(3s) | 0 | 0 | 0 | 0.1% | 0.3% | 0 | 0 | 0 | 0 | 0 |
| Cl(3p) | 11.8% | 8.4% | 8.8% | 3.2% | 7.9% | 2.8% | 7.8% | 6.2% | 4.8% | 4.6% |
| O(2s) | 0.3% | 2.2% | 0.3% | 8.8% | 7.8% | 7.2% | 0 | 4.2% | 1.4% | 6.6% |
| O(2p) | 0 | 0.1% | 1.7% | 5.4% | 1.8% | 1.8% | 1.5% | 3.9% | 0.2% | 3.5% |
| H(1s) | 2.0% | 4.9% | 0.7% | 37.7% | 23.1% | 22.5% | 0.9% | 21.6% | 6.8% | 28.3% |
| Total | 96.6% | 94.7% | 96.4% | 62.3% | 75.4% | 80.8% | 95.8% | 75.6% | 89.8% | 63.5% |

**Table S16.** Atomic orbital (AO) contributions to the five molecular states (i.e., the α-HOMO and four higher energy MOs) used for constructing real-space molecular orbital (MO) projectors for each spin for TiClMeCN (TiCl$_3$(CH$_3$CN)$_3$) in its ($n_α$, $n_β$) = (1, 0) $d^1$ doublet state along with the β LUMO and four higher energy β MOs. Contributions are summed by ligand element type but distinguished by principal quantum number and subshell. A sum less than 100% indicates a lack of suitable projection of the extended state onto the all-electron atomic states from the pseudopotentials.

| Atomic Orbitals | α | | | | | β | | | | |
|---|---|---|---|---|---|---|---|---|---|---|
| | HOMO | LUMO | LUMO+1 | LUMO+2 | LUMO+3 | LUMO | LUMO+1 | LUMO+2 | LUMO+3 | LUMO+4 |
| Ti(3s) | 0 | 0 | 0 | 0 | 0 | 0 | 0 | 0 | 0 | 0 |
| Ti(3p) | 0 | 0 | 0 | 0 | 0 | 0 | 0 | 0 | 0 | 0 |
| Ti(3d) | 64.5% | 69.7% | 71.0% | 0.1% | 1.3% | 65.9% | 46.8% | 66.4% | 0 | 1.5% |
| Ti(4s) | 0 | 0 | 0 | 0.7% | 0 | 0 | 0 | 0 | 0.7% | 0 |
| Cl(3s) | 0 | 0 | 0 | 0 | 0 | 0 | 0 | 0 | 0 | 0 |
| Cl(3p) | 11.1% | 12.6% | 13.9% | 0 | 0 | 10.6% | 4.8% | 11.6% | 0 | 0.4% |
| C(2s) | 0 | 0 | 0 | 9.9% | 0.6% | 0 | 0 | 0 | 10.2% | 0 |
| C(2p) | 15.7% | 10.4% | 7.7% | 1.7% | 39.2% | 13.6% | 27.6% | 11.0% | 0.1% | 40.1% |
| N(2s) | 0 | 0 | 0 | 0.3% | 0.2% | 0 | 0 | 0 | 0.2% | 0 |
| N(2p) | 3.8% | 3.2% | 3.2% | 2.2% | 33.5% | 4.8% | 11.5% | 5.4% | 0.3% | 34.6% |
| H(1s) | 2.5% | 1.5% | 1.4% | 16.8% | 13.0% | 2.4% | 5.3% | 2.2% | 16.9% | 13.6% |
| Total | 97.6% | 97.4% | 97.2% | 31.7% | 87.8% | 97.3% | 96.0% | 96.6% | 28.4% | 90.2% |

**Table S17.** Atomic orbital (AO) contributions to the five molecular states (i.e., the α-HOMO and four higher energy MOs) used for constructing real-space molecular orbital (MO) projectors for each spin for TiBrNMe (TiBr$_3$(N(CH$_3$)$_3$)$_2$) in its ($n_α$, $n_β$) = (1, 0) $d^1$ doublet state along with the β LUMO and four higher energy β MOs. Contributions are summed by ligand element type but distinguished by principal quantum number and subshell. A sum less than 100% indicates a lack of suitable projection of the extended state onto the all-electron atomic states from the pseudopotentials.

| Atomic Orbitals | α | | | | | β | | | | |
|---|---|---|---|---|---|---|---|---|---|---|
| | HOMO | LUMO | LUMO+1 | LUMO+2 | LUMO+3 | LUMO | LUMO+1 | LUMO+2 | LUMO+3 | LUMO+4 |
| Ti(3s) | 0 | 0 | 0 | 0 | 0 | 0 | 0 | 0 | 0 | 0 |
| Ti(3p) | 0 | 0 | 0 | 0 | 0 | 0 | 0 | 0 | 0 | 0 |
| Ti(3d) | 86.3% | 87.1% | 76.5% | 74.7% | 68.9% | 88.0% | 78.4% | 76.8% | 87.6% | 57.8% |
| Ti(4s) | 0 | 0 | 0 | 0 | 0.5% | 0 | 0 | 0 | 0 | 1.1% |
| Br(4s) | 0 | 0 | 0.5% | 0.4% | 0.6% | 0 | 0.4% | 0.5% | 0 | 0.7% |
| Br(4p) | 9.3% | 7.7% | 17.4% | 17.9% | 8.1% | 6.1% | 14.2% | 15.0% | 4.0% | 6.8% |
| C(2s) | 0 | 0 | 0 | 0 | 0 | 0 | 0 | 0 | 0.4% | 1.4% |
| C(2p) | 0.3% | 0.2% | 0 | 0 | 0.1% | 0.2% | 0 | 0 | 0.5% | 0.1% |
| N(2s) | 0 | 0 | 0 | 0 | 2.8% | 0 | 0 | 0 | 0 | 2.0% |
| N(2p) | 0 | 0 | 0 | 0 | 5.9% | 0 | 0 | 0 | 0 | 3.2% |
| H(1s) | 0.4% | 0.6% | 0 | 0 | 3.7% | 0.7% | 0 | 0 | 1.0% | 5.2% |
| Total | 96.3% | 95.6% | 94.4% | 93.0% | 90.6% | 95.0% | 93.0% | 92.3% | 93.5% | 78.3% |

**Table S18.** Atomic orbital (AO) contributions to the five molecular states (i.e., the α-HOMO and four higher energy MOs) used for constructing real-space molecular orbital (MO) projectors for each spin for TiAcac (Ti(acac)$_3$) in its ($n_\alpha$, $n_\beta$) = (1, 0) $d^1$ doublet state, where 'acac' corresponds to the acetylacetonate ligand, along with the β LUMO and four higher energy β MOs. Contributions are summed by ligand element type but distinguished by principal quantum number and subshell. A sum less than 100% indicates a lack of suitable projection of the extended state onto the all-electron atomic states from the pseudopotentials.

| Atomic Orbitals | α | | | | | β | | | | |
|---|---|---|---|---|---|---|---|---|---|---|
| | HOMO | LUMO | LUMO+1 | LUMO+2 | LUMO+3 | LUMO | LUMO+1 | LUMO+2 | LUMO+3 | LUMO+4 |
| Ti(3$s$) | 0 | 0 | 0 | 0 | 0 | 0 | 0 | 0 | 0 | 0 |
| Ti(3$p$) | 0 | 0 | 0 | 0 | 0 | 0 | 0 | 0 | 0 | 0 |
| Ti(3$d$) | 54.6% | 46.2% | 48.1% | 35.6% | 37.9% | 27.9% | 39.4% | 43.1% | 41.9% | 45.6% |
| Ti(4$s$) | 0 | 0 | 0 | 0 | 0 | 0 | 0 | 0 | 0 | 0 |
| C(2$s$) | 0 | 0 | 0 | 0 | 0 | 0 | 0 | 0 | 0 | 0 |
| C(2$p$) | 30.1% | 32.2% | 30.5% | 34.0% | 32.0% | 48.6% | 37.2% | 34.6% | 31.0% | 27.5% |
| O(2$s$) | 0 | 0 | 0 | 0 | 0 | 0 | 0 | 0 | 0 | 0 |
| O(2$p$) | 7.5% | 14.6% | 14.3% | 21.7% | 21.6% | 12.6% | 14.8% | 14.2% | 18.8% | 18.2% |
| H(1$s$) | 4.0% | 4.0% | 3.8% | 5.2% | 5.1% | 6.6% | 4.9% | 4.5% | 5.0% | 4.4% |
| Total | 96.2% | 97.0% | 96.7% | 96.5% | 96.6% | 95.7% | 96.3% | 96.4% | 96.7% | 95.7% |

**Table S19.** Atomic orbital (AO) contributions to the five molecular states (i.e., the α-HOMO and four higher energy MOs) used for constructing real-space molecular orbital (MO) projectors for each spin for VAcac (V(acac)$_3$) in its ($n_\alpha$, $n_\beta$) = (1, 0) $d^1$ doublet state, where 'acac' corresponds to the acetylacetonate ligand, along with the β LUMO and four higher energy β MOs. Contributions are summed by ligand element type but distinguished by principal quantum number and subshell. A sum less than 100% indicates a lack of suitable projection of the extended state onto the all-electron atomic states from the pseudopotentials.

| Atomic Orbitals | α | | | | | β | | | | |
|---|---|---|---|---|---|---|---|---|---|---|
| | HOMO | LUMO | LUMO+1 | LUMO+2 | LUMO+3 | LUMO | LUMO+1 | LUMO+2 | LUMO+3 | LUMO+4 |
| V(3$s$) | 0 | 0 | 0 | 0 | 0 | 0 | 0 | 0 | 0 | 0 |
| V(3$p$) | 0 | 0 | 0 | 0 | 0 | 0 | 0 | 0 | 0 | 0 |
| V(3$d$) | 75.3% | 71.7% | 66.2% | 9.6% | 9.2% | 79.1% | 73.4% | 61.3% | 5.5% | 5.6% |
| V(4$s$) | 0 | 0 | 0 | 0 | 0 | 0 | 0 | 0 | 0 | 0 |
| V(4$p$) | 0 | 0.2% | 0 | 0 | 0 | 0.2% | 0 | 0 | 0 | 0 |
| C(2$s$) | 0 | 0.1% | 0 | 0 | 0 | 0 | 0 | 0 | 0 | 0 |
| C(2$p$) | 8.9% | 7.9% | 10.8% | 54.1% | 52.6% | 7.0% | 9.6% | 27.2% | 58.1% | 57.4% |
| O(2$s$) | 0.6% | 0.3% | 0 | 0.7% | 0.4% | 0.4% | 0 | 0 | 0 | 0 |
| O(2$p$) | 11.6% | 17.4% | 20.6% | 24.3% | 25.5% | 10.8% | 14.0% | 5.8% | 24.4% | 26.0% |
| H(1$s$) | 1.1% | 0.2% | 0 | 7.4% | 7.7% | 0.1% | 0 | 3.6% | 8.4% | 8.2% |
| Total | 97.5% | 97.8% | 97.6% | 96.1% | 95.4% | 97.6% | 97.0% | 97.9% | 96.4% | 97.2% |

**Table S20.** Atomic orbital (AO) contributions to the five molecular states (i.e., the α-HOMO and four higher energy MOs) used for constructing real-space molecular orbital (MO) projectors for each spin for VOAcac (VO(acac)$_2$) in its ($n_α$, $n_β$) = (1, 0) $d^1$ doublet state, where 'acac' corresponds to the acetylacetonate ligand, along with the β LUMO and four higher energy β MOs. Contributions are summed by ligand element type but distinguished by principal quantum number and subshell. A sum less than 100% indicates a lack of suitable projection of the extended state onto the all-electron atomic states from the pseudopotentials.

| Atomic Orbitals | α | | | | | β | | | | |
|---|---|---|---|---|---|---|---|---|---|---|
| | HOMO | LUMO | LUMO+1 | LUMO+2 | LUMO+3 | LUMO | LUMO+1 | LUMO+2 | LUMO+3 | LUMO+4 |
| V(3$s$) | 0 | 0 | 0 | 0 | 0 | 0 | 0 | 0 | 0 | 0 |
| V(3$p$) | 0 | 0 | 0 | 0 | 0 | 0 | 0 | 0 | 0 | 0 |
| V(3$d$) | 86.0% | 8.3% | 6.1% | 63.3% | 57.3% | 5.3% | 4.0% | 90.1% | 65.3% | 62.9% |
| V(4$s$) | 0.1% | 0 | 0 | 0 | 0 | 0 | 0 | 0.4% | 0 | 0 |
| V(4$p$) | 0 | 0 | 0 | 0.3% | 0.4% | 0 | 0 | 0 | 0.5% | 0.5% |
| C(2$s$) | 0.1% | 0 | 0 | 0.1% | 0.2% | 0 | 0 | 0.4% | 0.5% | 0.5% |
| C(2$p$) | 0.3% | 54.5% | 57.3% | 2.3% | 5.9% | 57.6% | 58.8% | 0.2% | 3.0% | 2.9% |
| O(2$s$) | 0.8% | 0 | 0.7% | 1.0% | 1.9% | 0 | 0.2% | 0.4% | 1.2% | 1.9% |
| O(2$p$) | 10.2% | 26.5% | 24.8% | 28.8% | 28.5% | 25.7% | 25.3% | 5.8% | 22.9% | 23.5% |
| H(1$s$) | 0.2% | 7.8% | 8.4% | 0.1% | 1.5% | 8.6% | 8.8% | 0 | 0.6% | 1.3% |
| Total | 97.7% | 97.1% | 97.3% | 95.9% | 95.7% | 97.2% | 97.1% | 97.3% | 94.0% | 93.5% |

**Table S21.** Atomic orbital (AO) contributions to the five molecular states (i.e., the α-HOMO and four higher energy MOs) used for constructing real-space molecular orbital (MO) projectors for each spin for VOPrAq (VO(prz)$_2$(H$_2$O)) in its ($n_α$, $n_β$) = (1, 0) $d^1$ doublet state, where 'prz' corresponds to the pyrazinoate ligand, along with the β LUMO and four higher energy β MOs. Contributions are summed by ligand element type but distinguished by principal quantum number and subshell. A sum less than 100% indicates a lack of suitable projection of the extended state onto the all-electron atomic states from the pseudopotentials.

| Atomic Orbitals | α | | | | | β | | | | |
|---|---|---|---|---|---|---|---|---|---|---|
| | HOMO | LUMO | LUMO+1 | LUMO+2 | LUMO+3 | LUMO | LUMO+1 | LUMO+2 | LUMO+3 | LUMO+4 |
| V(3$s$) | 0 | 0 | 0 | 0 | 0 | 0 | 0 | 0 | 0 | 0 |
| V(3$p$) | 0 | 0 | 0 | 0 | 0 | 0 | 0 | 0 | 0 | 0 |
| V(3$d$) | 75.4% | 6.5% | 9.8% | 19.2% | 6.1% | 20.7% | 7.2% | 6.8% | 28.8% | 37.2% |
| V(4$s$) | 0 | 0 | 0 | 0 | 0 | 0 | 0 | 0 | 0 | 0 |
| V(4$p$) | 0.2% | 0.2% | 0.1% | 0.4% | 0 | 0.2% | 0 | 0.1% | 0 | 0 |
| C(2$s$) | 0 | 0 | 0 | 0 | 0 | 0 | 0 | 0 | 0 | 0.2% |
| C(2$p$) | 7.6% | 36.0% | 34.6% | 60.7% | 77.6% | 32.7% | 36.8% | 79.6% | 55.7% | 41.5% |
| N(2$s$) | 0 | 0 | 0 | 0 | 0 | 0 | 0 | 0 | 0 | 0 |
| N(2$p$) | 4.7% | 44.5% | 45.7% | 2.4% | 4.5% | 37.0% | 46.3% | 0.7% | 3.8% | 14.7% |
| O(2$s$) | 0.2% | 0 | 0.2% | 0.1% | 0 | 0 | 0 | 0.1% | 0.2% | 0.2% |
| O(2$p$) | 9.6% | 9.9% | 6.2% | 14.0% | 7.6% | 7.2% | 7.0% | 8.9% | 8.7% | 2.5% |
| H(1$s$) | 0.2% | 0 | 0.2% | 0 | 0 | 0 | 0 | 0 | 0.2% | 0.4% |
| Total | 97.9% | 97.1% | 96.8% | 96.8% | 95.8% | 97.8% | 97.3% | 96.2% | 97.4% | 96.7% |

**Table S22.** Atomic orbital (AO) contributions to the five molecular states (i.e., the α-HOMO and four higher energy MOs) used for constructing real-space molecular orbital (MO) projectors for each spin for VOClNMe (VOCl$_2$(NH(CH$_3$)$_2$)$_2$) in its ($n_α$, $n_β$) = (1, 0) $d^1$ doublet state, along with the β LUMO and four higher energy β MOs. Contributions are summed by ligand element type but distinguished by principal quantum number and subshell. A sum less than 100% indicates a lack of suitable projection of the extended state onto the all-electron atomic states from the pseudopotentials.

| Atomic Orbitals | α | | | | | β | | | | |
|---|---|---|---|---|---|---|---|---|---|---|
| | HOMO | LUMO | LUMO+1 | LUMO+2 | LUMO+3 | LUMO | LUMO+1 | LUMO+2 | LUMO+3 | LUMO+4 |
| V(3$s$) | 0 | 0 | 0 | 0 | 0 | 0 | 0 | 0 | 0 | 0 |
| V(3$p$) | 0 | 0 | 0 | 0.1% | 0.1% | 0 | 0 | 0.1% | 0 | 0 |
| V(3$d$) | 82.6% | 66.7% | 62.0% | 63.2% | 32.6% | 88.0% | 71.3% | 64.3% | 66.6% | 2.6% |
| V(4$s$) | 0 | 0 | 0 | 0.2% | 0.6% | 0 | 0 | 0 | 0.3% | 4.6% |
| V(4$p$) | 0 | 0.7% | 1.0% | 0.8% | 1.6% | 0 | 0.8% | 2.0% | 0.3% | 0 |
| Cl(3$s$) | 0 | 0 | 0.9% | 1.0% | 0.4% | 0 | 0 | 0.8% | 1.1% | 0.6% |
| Cl(3$p$) | 14.6% | 1.6% | 11.1% | 15.2% | 2.1% | 7.0% | 1.4% | 11.8% | 11.3% | 0.5% |
| C(2$s$) | 0 | 0 | 0 | 0 | 3.0% | 0.1% | 0.2% | 0 | 0 | 5.7% |
| C(2$p$) | 0 | 0.7% | 0 | 0 | 0 | 0 | 0.7% | 0 | 0 | 1.7% |
| N(2$s$) | 0 | 0.8% | 0.2% | 1.2% | 0.8% | 0 | 0.8% | 0.7% | 0.5% | 1.2% |
| N(2$p$) | 0 | 1.9% | 0.1% | 3.0% | 4.2% | 0 | 1.4% | 1.2% | 1.3% | 1.4% |
| O(2$s$) | 0 | 0 | 0.1% | 0.2% | 1.1% | 0 | 0 | 0.3% | 0 | 0.5% |
| O(2$p$) | 0 | 25.1% | 20.2% | 9.8% | 9.8% | 0 | 20.3% | 12.5% | 13.6% | 0.4% |
| H(1$s$) | 1.0% | 0 | 0 | 0.4% | 9.0% | 1.7% | 0.3% | 0.1% | 0.2% | 16.1% |
| Total | 98.2% | 97.5% | 95.6% | 95.0% | 65.2% | 96.8% | 97.2% | 93.7% | 95.2% | 35.3% |

**Table S23.** Atomic orbital (AO) contributions to the five molecular states (i.e., the α-HOMO and four higher energy MOs) used for constructing real-space molecular orbital (MO) projectors for each spin for VClNCH (VCl$_4$(NCH)$_2$) in its ($n_α$, $n_β$) = (1, 0) $d^1$ doublet state, along with the β LUMO and four higher energy β MOs. Contributions are summed by ligand element type but distinguished by principal quantum number and subshell. A sum less than 100% indicates a lack of suitable projection of the extended state onto the all-electron atomic states from the pseudopotentials.

| Atomic Orbitals | α | | | | | β | | | | |
|---|---|---|---|---|---|---|---|---|---|---|
| | HOMO | LUMO | LUMO+1 | LUMO+2 | LUMO+3 | LUMO | LUMO+1 | LUMO+2 | LUMO+3 | LUMO+4 |
| V(3$s$) | 0 | 0 | 0 | 0 | 0 | 0 | 0 | 0 | 0 | 0 |
| V(3$p$) | 0 | 0 | 0 | 0 | 0 | 0 | 0 | 0 | 0 | 0 |
| V(3$d$) | 65.0% | 68.7% | 67.7% | 56.4% | 55.8% | 72.1% | 72.1% | 63.8% | 46.0% | 57.1% |
| V(4$s$) | 0 | 0 | 0 | 0 | 0 | 0 | 0 | 0 | 0 | 0 |
| V(4$p$) | 0 | 0 | 0 | 0 | 0 | 0 | 0 | 0 | 0 | 0 |
| Cl(3$s$) | 0 | 0 | 0.3% | 2.2% | 2.3% | 0 | 0.3% | 0 | 1.3% | 2.7% |
| Cl(3$p$) | 29.6% | 28.3% | 30.1% | 26.3% | 26.6% | 22.7% | 24.4% | 15.5% | 14.3% | 26.7% |
| C(2$s$) | 0 | 0 | 0 | 0.4% | 0.4% | 0 | 0 | 0 | 0.6% | 0.3% |
| C(2$p$) | 3.4% | 1.3% | 0.3% | 3.6% | 3.7% | 2.6% | 1.4% | 13.5% | 16.1% | 2.9% |
| N(2$s$) | 0.4% | 0 | 0 | 3.6% | 3.5% | 0 | 0 | 0.2% | 3.8% | 2.2% |
| N(2$p$) | 0.4% | 0.1% | 0 | 3.8% | 3.9% | 0.6% | 0 | 4.7% | 12.4% | 3.0% |
| H(1$s$) | 0 | 0 | 0 | 0.6% | 0.6% | 0 | 0 | 0 | 0.8% | 0.5% |
| Total | 98.8% | 98.4% | 98.4% | 96.9% | 96.8% | 98.0% | 98.2% | 97.7% | 95.3% | 95.4% |

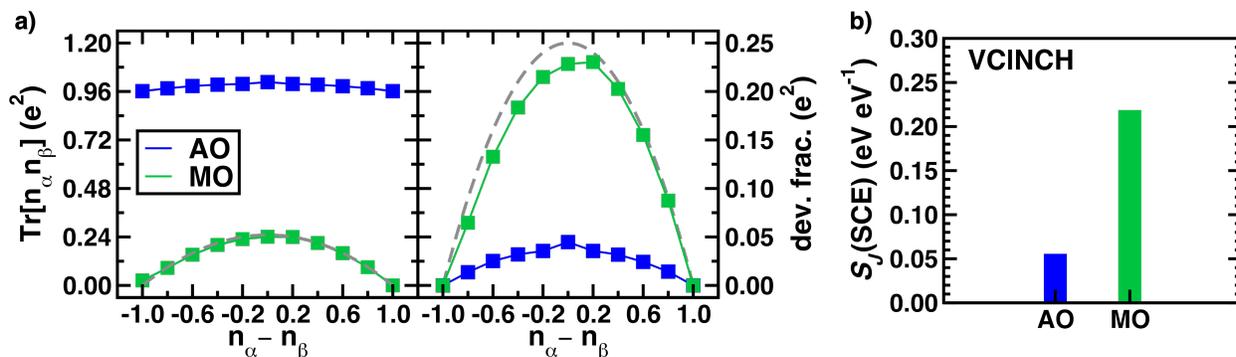

**Figure S2.** a) PBE GGA fractionalities (i.e., $\text{Tr}[\mathbf{n}_\alpha \mathbf{n}_\beta]$ in e$^{-2}$, left) and deviations from linear admixture (dev. frac. in e$^{-2}$, right) for VClNCH (i.e., VCl$_4$(NCH)$_2$) along the FSL (i.e., where the $d^1$ doublet state switches from an ($n_\alpha$, $n_\beta$) = (0, 1) configuration to an ($n_\alpha$, $n_\beta$) = (1, 0) configuration). The fractionalities and deviations were obtained using both atomic orbital (AO) projectors (blue squares) and molecular orbital (MO) projectors obtained from the state with formal $d^1$ configuration (green squares). Fractionalities and fractionality deviations at the ideal atomic limit are indicated using gray dashed lines. b) Linearized sensitivities to the variation in static correlation error (SCE) with increasingly negative $J$ values at $U = 0$ ($S_J$(SCE), in eV/eV of $J$) evaluated for VClNCH using AO projectors (blue bar) and MO projectors (green bar).

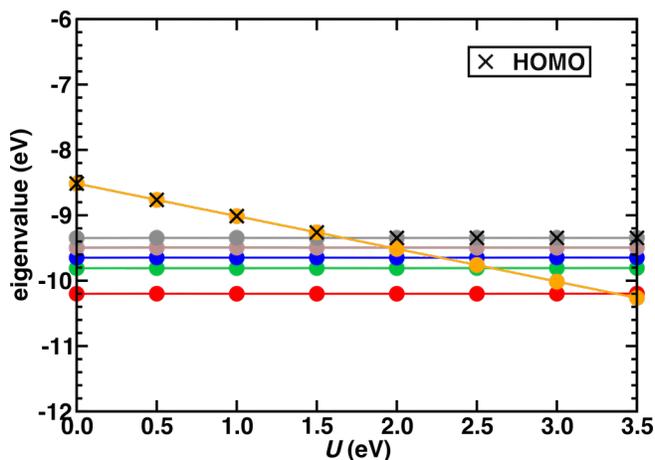

**Figure S3.** Frontier majority-spin (α) orbital eigenvalue variation with an increase in $U$ value at $J = 0$ eV for [V(acac)$_3$]$^+$, where V is in a formal $d^1$ electron configuration, computed using MO projectors. Highest occupied molecular orbital (HOMO) at each $U$ value is marked with a black cross. From $U = 2.0$ eV onwards, the MO that overlaps the most with the selected MO projectors (orange) is no longer the HOMO. Therefore, the HOMO eigenvalue levels off leading to a reduced EDE reduction efficiency with the MO-DFT+U correction.

**Table S24.** Energetic delocalization error (EDE) and static correlation error (SCE) computed using the absolute maximum deviation (in eV) from linearity and energetic invariance occurring at the midpoint of FCL and FSL respectively. All errors are computed using the PBE functional and with the non-empirical jmDFT correction using standard metal-centered $d$ atomic orbital projectors (AOP) and using molecular orbital projectors (MOP). The chemical formula of each complex as well as the nomenclature assigned to it throughout this work are both indicated. In the listed chemical formulae, 'acac' corresponds to the acetylacetonate ligand while 'prz' corresponds to the pyrazinoate ligand.

| Nomenclature | Chemical Formula | EDE (eV) | | | SCE (eV) | | |
|---|---|---|---|---|---|---|---|
| | | PBE | AOP | MOP | PBE | AOP | MOP |
| TiClAq | $TiCl_3(H_2O)_3$ | 0.70 | 0.34 | 0.01 | 0.19 | 0.12 | 0.05 |
| TiClMeCN | $TiCl_3(CH_3CN)_3$ | 0.58 | 0.21 | 0.07 | 0.22 | 0.17 | 0.08 |
| TiBrNMe | $TiBr_3(N(CH_3)_3)_2$ | 0.60 | 0.35 | 0.04 | 0.21 | 0.12 | 0.21 |
| TiAcac | $Ti(acac)_3$ | 0.52 | 0.30 | 0.09 | 0.20 | 0.20 | 0.00 |
| VAcac | $V(acac)_3$ | 0.51 | 0.42 | 0.16 | 0.29 | 0.17 | 0.08 |
| VOAcac | $VO(acac)_2$ | 0.69 | 0.48 | 0.04 | 0.50 | 0.24 | 0.15 |
| VOPrAq | $VO(prz)_2(H_2O)$ | 0.58 | 0.37 | 0.16 | 0.27 | 0.17 | 0.05 |
| VOClNMe | $VOCl_2(NH(CH_3)_2)_2$ | 0.68 | 0.34 | 0.07 | 0.46 | 0.12 | 0.05 |
| VClNCH | $VCl_4(NCH)_2$ | 0.63 | 0.48 | 0.10 | 0.20 | 0.15 | 0.04 |

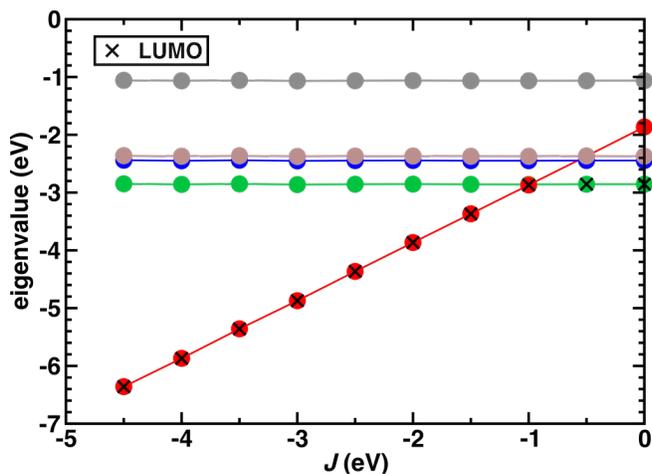

**Figure S4.** Frontier minority-spin (β) molecular orbital eigenvalue variation with an increase in $J$ value at $U = 0$ eV for $TiBr_3(N(CH_3)_3)_2$, where Ti is in a formal $d^1$ electron configuration, computed using MO projectors. Lowest unoccupied molecular orbital (LUMO) at each $J$ value is marked with a black cross. The LUMO eigenvalue only starts varying below $J = -1$ eV. At the non-empirical jmDFT $U + J$ value of -0.84 eV, MO jmDFT with non-empirical values do not shift the SCE.